\documentclass[aps,superscriptaddress,twocolumn,showpacs,dvips]{revtex4}
\usepackage{feynmp}
\usepackage{amssymb}
\usepackage{epsfig}
\usepackage{graphicx}
\usepackage{subfigure}
\usepackage{hyperref}
\usepackage{bbm}
\usepackage{gensymb}
\usepackage{float}
\usepackage{xcolor}

\begin{document}

\title{Superconductivity in two-dimensional disordered Dirac semimetals}

\author{Jing Wang}
\affiliation{Department of Modern Physics, University of Science and
Technology of China, Hefei, Anhui 230026, P. R. China}
\author{Peng-Lu Zhao}
\affiliation{Department of Modern Physics, University of Science and
Technology of China, Hefei, Anhui 230026, P. R. China}
\author{Jing-Rong Wang}
\affiliation{High Magnetic Field Laboratory, Hefei Institutes of
Physical Science, Chinese Academy of Sciences, Hefei, Anhui 230031,
P. R. China}
\author{Guo-Zhu Liu}
\altaffiliation{Corresponding author: gzliu@ustc.edu.cn}
\affiliation{Department of Modern Physics, University of Science and
Technology of China, Hefei, Anhui 230026, P. R. China}

\begin{abstract}
In two-dimensional Dirac semimetals, Cooper pairing instability
occurs only when the attractive interaction strength $|u|$ is larger
than some critical value $|u_{c}|$ because the density of states
vanishes at Dirac points. Disorders enhance the low-energy density
of states but meanwhile shorten the lifetime of fermions, which tend
to promote and suppress superconductivity, respectively. To
determine which of the two competing effects wins, we study the
interplay of Cooper pairing interaction and disorder scattering by
means of renormalization group method. We consider three types of
disorders, including random mass, random gauge potential, and random
chemical potential, and show that the first two suppress
superconductivity. In particular, the critical BCS coupling
$|u_{c}|$ is increased to certain larger value if the system
contains only random mass or random gauge potential, which makes the
onset of superconductivity more difficult. In the case of random
chemical potential, the effective disorder parameter flows to the
strong coupling regime, where the perturbation expansion breaks down
and cannot provide a clear answer concerning the fate of
superconductivity. When different types of disorder coexist in one
system, their strength parameters all flow to strong couplings. In
the strong coupling regime, the perturbative renormalization group
method becomes invalid, and one needs to employ other methods to
treat the disorder effects. We perform a simple gap equation
analysis of the impact of random chemical potential on
superconductivity by using the Abrikosov-Gorkov diagrammatic
approach, and also briefly discuss the possible generalization of
this approach.
\end{abstract}


\maketitle


\section{Introduction}

In the Bardeen-Cooper-Schrieffer (BCS) theory of metal
superconductors, the Cooper pairing instability caused by a net
attractive interaction plays an essential role. A pair of electrons
can be bound together by an arbitrarily weak attractive force
between them, known as the Cooper theorem. This theorem can by
reformulated in the modern renormalization group (RG) theory, which
states that the attractive interaction, characterized by a negative
coupling constant $u$, is a (marginally) relevant perturbation to
the electronic system \cite{Shankar1994RMP}. The RG equation for $u$
takes the general form
\begin{eqnarray}
\frac{du}{dl} = -c u^2 \label{Eq_u_c}
\end{eqnarray}
in three-dimensional (3D) metals, where $l$ is a varying length
scale and $c$ some constant. This equation tells us that, while a
positive $u$ would flow to zero at large $l$, a negative $u$ flows
indefinitely to the strong coupling regime no matter how small its
initial value is.

In the past three decades, there have been a great deal of research
activities devoted to studying the physical properties of electronic
systems in which the valence and conduction bands touch at a number
of discrete points. Examples include zero-gap semiconductors
\cite{Fradkin1986}, $d$-wave high-$T_c$ cuprate superconductors
\cite{Orenstein, Lee2006}, graphene \cite{CastroNeto, Kotov,
DasSarma}, topological insulators \cite{HasanKane, ZhangQi, Vafek},
and Weyl semimetals \cite{Vafek}. These materials exhibit different
low-energy behaviors. However, irrespective of the microscopic
details, a very common feature shared by these materials is that
their low-energy fermionic excitations have a linear dispersion and
thus can be described by $N$ species of massless Dirac fermions.
Extensive recent theoretic studies \cite{Orenstein, Lee2006, Vafek,
CastroNeto, Kotov, DasSarma} have elaborated that the interparticle
interactions can result in non-Fermi-liquid behaviors and certain
quantum phase transition. For instance, the long-range Coulomb
interaction is able to drive an excitonic insulating transition if
its strength is sufficiently large \cite{Khveshchenko01, Gorbar02,
Khveshchenko04, Liu09, LiuNJP11, WangNJP12}. The Coulomb interaction
can also give rise to unusual spectral behaviors of Dirac fermions
\cite{CastroNeto, Kotov}. Moreover, the strong spin-orbit coupling
may open a finite gap and as such produce a topological insulator
\cite{Vafek, HasanKane}.

The Cooper pairing of Dirac fermions and the resultant
superconducting transition are two subjects of considerable interest
\cite{Uchoa05, Uchoa07, Kopnin08, Gonzalez08, Zhao06, Honerkamp08,
Roy10, Roy13, Roy14, Lee14, Yao15, Maciejko16, Chubukov12, Sondhi13,
Sondhi14}. The superconductivity might be mediated by various
bosonic modes, such as phonons or plasmons, in graphene
\cite{Uchoa07}. When graphene is properly doped such that its Fermi
surface is close to a Van Hove singularity \cite{Gonzalez08,
Chubukov12}, even repulsive interaction can result in
superconductivity via the Kohn-Luttinger mechanism \cite{Kohn65}.
Other pairing mechanisms are also possible and have been extensively
studied \cite{Roy13}. Recent experiment has revealed direct evidence
for emergent superconductivity on the surface of a 3D topological
insulator \cite{Krusin15}.

The Cooper pairing in intrinsic two-dimensional (2D) Dirac
semimetals in which the Fermi energy is tuned to precisely the Dirac
points is particularly interesting. Previous theoretical and
numerical studies \cite{Uchoa07, Zhao06, Honerkamp08, Sondhi13} have
found that, different from ordinary metals, the Cooper theorem is no
longer valid in undoped Dirac semimetals. Because the fermion
density of states (DOS) vanishes linearly near the band-touching
points, an infinitely weak attraction is not sufficient to bind
Dirac fermions together to form Cooper pairs. Cooper pairing can be
triggered only when the attraction strength exceeds certain
threshold \cite{Uchoa07, Zhao06, Honerkamp08, Sondhi13}, i.e., $|u|
> |u_c|$ with $u_c$ being a finite critical value. Hence, there is a
quantum phase transition between the semimetallic and
superconducting phases, and the critical value $u_c$ defines the
quantum critical point (QCP). It was proposed recently that a
space-time supersymmetry emerges at such a quantum critical point
\cite{Lee14, Maciejko16}, where the massless Dirac fermions and the
massless bosonic order parameter are connected by a superconformal
algebra.

An interesting question is whether the semimetal-superconductor
quantum critical point and the emergent supersymmetry are robust
against disorders. To answer this question, it is necessary to
investigate the impact of disorders on Cooper pairing. It seems that
disorders can promote superconductivity since it may generate a
finite zero-energy DOS. However, this could happen only in the
presence of disorder that is a relevant perturbation to the system.
On the other hand, disorder scattering could break Cooper pairs by
shortening the lifetime of Dirac fermions, which would destruct
superconductivity. Moreover, there are at least three types of
disorder in Dirac semimetals \cite{Ludwig1994, Nersesyan95,
Altland02}, including random chemical potential, random mass, and
random gauge potential. They have various physical origins, couple
to Dirac fermions in different manners, and can result in distinct
low-energy behaviors \cite{Ludwig1994, Nersesyan95, Altland02,
Stauber2005PRB, Wang2011PRB, WangLiu2014}. It thus turns out that
the effects of disorders are rich and complicated. To acquire a
clear understanding of disorder effects, one needs to treat the
quartic pairing interaction and fermion-disorder coupling on an
equal footing, and analyze how the critical coupling $u_c$ is
affected by various types of disorder.

Motivated by the above consideration, we will investigate in this
paper the disorder effects by performing a perturbative RG analysis
within an effective model that contains both quartic pairing
interaction and fermion-disorder coupling. Recently, Nandkishore
\emph{et al.} \cite{Sondhi13} and Potirniche \emph{et al.}
\cite{Sondhi14} have studied the effects of random chemical
potential on superconductivity. The main conclusion reached in the
mean-field analysis \cite{Sondhi13} is that superconductivity is
enhanced. In this paper, we will consider the impact of all the
three types of disorder.

When the Dirac semimetal contains weak random mass or random gauge
potential, the time-reversal symmetry is broken and the Anderson
theorem \cite{Anderson1959JPCS, Gorkov08} is certainly invalid. We
study the fate of superconductivity by carrying out RG calculations
and find that, the critical BCS coupling $|u_{c}|$ increases to
certain larger value $|u_{c}'|$ when random mass or random gauge
potential exists by itself, which makes it more difficult to realize
Cooper pairing in realistic materials. The effective strength of
these two types of disorder either flows rapidly to zero or remains
a small constant at low energies, thus the perturbative RG expansion
is under control and the RG results are reliable. We therefore can
conclude that superconductivity is more or less suppressed. If $|u|
< |u_{c}'|$, the Dirac semimetal remains gapless, but its low-energy
properties are strongly affected by disorder. Specifically, random
mass leads to marginal Fermi liquid (MFL-) like behavior, and random
gauge potential induces non-Fermi liquid (NFL) behavior. As
$|u_{0}|$ grows upon approaching $|u_{c}'|$, the Dirac semimetal
enters into a superconducting phase. The nature of such a QCP
depends sensitively on the specific type of disorder: for random
mass, $|u_{c}'|$ defines a QCP between a MFL-like phase and a
superconducting state; for random gauge potential, $|u_{c}'|$
defines a QCP between a NFL and a superconducting state.

If only random chemical potential exists, the effective strength
increases monotonously as the energy is lowered, which means that
the perturbative RG method is out of control in the low-energy
region and does not give us a clear answer to the fate of
superconductivity. Other efficient theoretic tool is needed to study
the effects of random chemical potential on superconductivity.

It is widely believed that random chemical potential generates a
finite zero-energy DOS, namely $\rho(0)\neq 0$. Similar to
Nandkishore \emph{et al.} \cite{Sondhi13}, we assume that the Dirac
fermion has only one flavor, with the surface state of a 3D
topological insulator being an example. In this case, there is no
conventional Anderson localization \cite{Ludwig1994, Mirlin2006PRB,
Evers08, Nomura07, Ryu10}, and the fermions become diffusive due to
random chemical potential, but remain extended. As this problem
cannot be handled by perturbative RG, one might appeal to the
mean-field analysis, such as the Abrikosov-Gorkov (AG) approach
\cite{Gorkov08, Sondhi13}. We will present a simple AG analysis and
derive the superconducting gap equation after incorporating the
impact of random chemical potential. However, it is important to
remember that the original AG approach entirely ignores vertex
corrections and is justified only in the limit $k_F \lambda \gg 1$,
where $k_F$ is the Fermi momentum and $\lambda$ the mean free path.
In 2D Dirac semimetal, we know that $k_F \rightarrow 0$, thus the
applicability of the AG approach is indeed not well justified. The
importance of the vertex corrections needs to be carefully examined,
which is an interesting task but goes beyond the scope of the
present paper.

After investigating the impact of each single type of disorder, we
also consider the coexistence of different types of disorder and
find that they have significant mutual influence on each other.
Actually, once more than one types of disorder exist in the system,
all three types of disorder are present and flow to strong couplings
at low energies, driving the system entering into a highly
disordered phase. In that case, the fate of superconductivity
remains undetermined.

The rest of the paper will be organized as follows. We present the
model Hamiltonian in Sec.~\ref{Sec_model} and studied the clean
limit in Sec.~\ref{Sec_clean}. We perform the detailed RG
calculations in Sec.~\ref{Sec_one_loop}, and then use the RG results
to determine the impact of disorder on superconductivity in Sec.~\ref{Sec_one_disorder}. 
The mutual influence between different disorders is also studied in this
section. We discuss the applicability of the AG approach in
Sec.~\ref{Sec_discuss}. We briefly summarize the results and
highlight further works in Sec.~\ref{Sec_summary}.

\section{Effective model}\label{Sec_model}

We begin with the following model Hamiltonian \cite{Sondhi13}:
\begin{eqnarray}
H = H_{0} + H_{\mathrm{int}}+H_{\mathrm{dis}},
\end{eqnarray}
which may describe the Dirac fermions on a 2D honeycomb lattice or
on the surface of a three-dimensional topological insulator. The
free term of Dirac fermions is
\begin{eqnarray}
H_{0} = \sum_{\mathbf{k}}\Psi^\dagger(\mathbf{k})(v_F k_x\sigma_1 +
v_F k_y\sigma_2 - \mu\sigma_0)\Psi(\mathbf{k}),
\end{eqnarray}
where $v_F$ is the Fermi velocity and $\mu$ chemical potential. We
use $\sigma_0$ to denote the $2 \times 2$ identity matrix, and
$\sigma_i$ with $i=1,2,3$ to denote the Pauli matrices, which
satisfy the algebra $\left\{\sigma_{i},\sigma_{j}\right\} =
2\delta_{ij}$. Since the goal of the present work is to examine the
possibility of superconductivity in intrinsic Dirac semimetals, we
assume that the Fermi surface is tuned to be exactly at the Dirac
points, and henceforth set $\mu = 0$. Moreover, we assume there is
one flavor of fermion, and neglect the possibility of Anderson
localization \cite{Ludwig1994, Evers08, Mirlin2006PRB, Nomura07, Ryu10}.

A possible quartic short-range interaction of Dirac fermions has the
following form
\begin{eqnarray}
H_{\mathrm{int}} = \int d^2\mathbf{x} \frac{u(\mathbf{x})}{4}
\Psi^\dagger(\mathbf{x})\sigma_0 \Psi(\mathbf{x})
\Psi^\dagger(\mathbf{x})\sigma_0 \Psi(\mathbf{x}).
\end{eqnarray}
For simplicity, the coupling function $u(\mathbf{x})$ can be
replaced by a constant $u$, which after renormalization will depend
on the varying energy scale. Making a Fourier transformation leads
to
\begin{eqnarray}
H_{\mathrm{int}} &=& \frac{u}{4}\int
\frac{d^{2}\mathbf{k}_1}{(2\pi)^2}\frac{d^{2}\mathbf{k}_2}{(2\pi)^2}
\frac{d^{2}\mathbf{k}_3}{(2\pi)^2} \nonumber \\
&& \times \Psi^\dagger_{\mathbf{k}_1}\sigma_0
\Psi_{\mathbf{k}_2}\Psi^\dagger_{\mathbf{k}_3}\sigma_0
\Psi_{\mathbf{k}_1 + \mathbf{k}_3 - \mathbf{k}_2},\label{Eq_H_int}
\end{eqnarray}
where the spinor $\Psi^\dagger_{\mathbf{k}} =
\left(c^\dagger_{\mathbf{k}\uparrow},
c^\dagger_{\mathbf{k}\downarrow}\right)$ and $\Psi^T_{\mathbf{k}}=
\left(c_{\mathbf{k}\uparrow}, c_{\mathbf{k}\downarrow}\right)$ are
introduced to describe Dirac fermions. Now we can expand the quartic
coupling term as
\begin{eqnarray}
&&\Psi^\dagger_{\mathbf{k}_1}(\sigma_0)\Psi_{\mathbf{k}_2}
\Psi^\dagger_{\mathbf{k}_3}(\sigma_0)\Psi_{\mathbf{k}_4} \nonumber\\
&=& c^\dagger_{\mathbf{k}_1\uparrow}c_{\mathbf{k}_2\uparrow}
c^\dagger_{\mathbf{k}_3\uparrow}c_{\mathbf{k}_4\uparrow}
+c^\dagger_{\mathbf{k}_1\uparrow}c_{\mathbf{k}_2\uparrow}
c^\dagger_{\mathbf{k}_3\downarrow}c_{\mathbf{k}_4\downarrow}\nonumber\\
&&+c^\dagger_{\mathbf{k}_1\downarrow}c_{\mathbf{k}_2\downarrow}
c^\dagger_{\mathbf{k}_3\uparrow}c_{\mathbf{k}_4\uparrow}
+c^\dagger_{\mathbf{k}_1\downarrow}c_{\mathbf{k}_2\downarrow}
c^\dagger_{\mathbf{k}_3\downarrow}c_{\mathbf{k}_4\downarrow},
\end{eqnarray}
with $\mathbf{k}_4=\mathbf{k}_1 + \mathbf{k}_3 - \mathbf{k}_2$. The
first and fourth terms involve spinors with the same spin if we
start from the interaction in Eq.~(\ref{Eq_H_int}), which are indeed
not allowed by the Pauli principle \cite{Sondhi13}. This implies
that the interaction can not capture all the potential four-fermion
interactions in a 2D Dirac semimetal. To remedy this, we follow the
approach of Ref.~\cite{Sondhi13} and consider another quartic
coupling term
\begin{eqnarray}
H_{\mathrm{int}}\sim\Psi^\dagger_{\mathbf{k}}(-i\sigma_2)
\Psi_{\mathbf{k+q}}\Psi^\dagger_{\mathbf{p}}(i\sigma_2)
\Psi_{\mathbf{p-q}},\label{Eq_H_int_eff}
\end{eqnarray}
which can be expanded to give
\begin{eqnarray}
&&\Psi^\dagger_{\mathbf{k}_1}(-i\sigma_2)\Psi_{\mathbf{k}_2}
\Psi^\dagger_{\mathbf{k}_3}(i\sigma_2)\Psi_{\mathbf{k}_4}\nonumber\\
&=&-c^\dagger_{\mathbf{k}_1\downarrow}c_{\mathbf{k}_2\uparrow}
c^\dagger_{\mathbf{k}_3\downarrow}c_{\mathbf{k}_4\uparrow}
+c^\dagger_{\mathbf{k}_1\downarrow}c_{\mathbf{k}_2\uparrow}
c^\dagger_{\mathbf{k}_3\uparrow}c_{\mathbf{k}_4\downarrow}\nonumber \\
&&+c^\dagger_{\mathbf{k}_1\uparrow}c_{\mathbf{k}_2\downarrow}
c^\dagger_{\mathbf{k}_3\downarrow}c_{\mathbf{k}_4\uparrow}
-c^\dagger_{\mathbf{k}_1\uparrow}c_{\mathbf{k}_2\downarrow}
c^\dagger_{\mathbf{k}_3\uparrow}c_{\mathbf{k}_4\downarrow},
\end{eqnarray}
which contains all four types of four-fermion coupling term and
hence can serve as the starting point.

\begin{figure}[htbp]
\center
\includegraphics[width=2.6in]{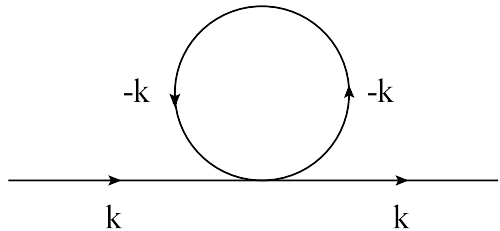}
\vspace{-0.2cm}\caption{One-loop correction to fermion propagator.
The solid line represents the free fermion propagator.}
\label{Fig_fermion_one_loop}
\end{figure}

We then consider the coupling between fermions and disorders, which
can be generically described by \cite{Ludwig1994, Nersesyan95,
Altland02, Stauber2005PRB, Wang2011PRB, WangLiu2014}
\begin{eqnarray}
H_{\mathrm{dis}} = v_{\Gamma}\int d^2\mathbf{x}
\Psi^\dagger(\mathbf{x})\Gamma\Psi(\mathbf{x})
A(\mathbf{x}),\label{Eq_H_dis}
\end{eqnarray}
where $v_{\Gamma}$ is a constant and the random field
$A(\mathbf{x})$ is taken to be a quenched, Gaussian variable
satisfying
\begin{eqnarray}
\langle A(\mathbf{x})\rangle = 0,\qquad \langle
A(\mathbf{x})A(\mathbf{x}')\rangle = \Delta
\delta^2(\mathbf{x}-\mathbf{x}')
\end{eqnarray}
with $\Delta$ being a dimensionless variance. The disorders are
classified by the definitions of the matrix $\Gamma$. More
concretely, $\Gamma = \sigma_0$ for random chemical (scalar)
potential, and $\Gamma = \sigma_3$ for random mass. In the case of
random gauge (vector) potential, there are two components for
$\Gamma$ and $v_{\Gamma}$: $\Gamma = (\sigma_{1},\sigma_{2})$ and
$v_{\Gamma} = (v_{\Gamma1},v_{\Gamma2})$. These disorders can be
induced by various mechanisms in realistic Dirac fermion materials.
For instance, the dominant impurity in $d$-wave cuprate
superconductors behaves like a random gauge potential
\cite{Nersesyan95}, whereas random mass and random chemical
potential appear in a 2D orbit antiferromagnet \cite{Nersesyan95}.
In the context of graphene, random chemical potential might be
produced by local defects or neutral absorbed atoms
\cite{Peres2010RMP, Mucciolo2010JPCM}. The ripple configurations of
graphene are usually described by a random gauge potential
\cite{CastroNeto, Meyer2007Nature}, and the random configurations in
the substrates can generate random mass \cite{Champel2010PRB,
Kusminskiy2011PRB}.

To make our consideration more generic, we suppose all the three
types of disorder coexist in the Dirac fermion system and derive the
RG equations for all the involved model parameters by employing the
replica method to average over the disordered potentials
\cite{Edwards1975, Lerner, Herbut2008PRL, Goswami2011PRL,
Roy2014PRB, Wang2015PLA, Roy2016PRB, Roy2016SCP}. The impact of each
single disorder on the fate of superconductivity can be readily
studied by removing the rest two types of disorder. We then consider
the interplay of different disorders and examine their mutual
influence.

There are three independent parameters in the total Hamiltonian: the
fermion velocity $v_F$, the quartic coupling constant $u$, and the
fermion-disorder coupling $\Delta_{\Gamma}$. They all flow under
scaling transformations, and might affect each other since the flow
equations are coupled. We will adopt the momentum-shell scheme of RG
approach \cite{Shankar1994RMP}, so it is most convenient to rewrite
the effective action in the momentum space
\begin{widetext}
\begin{eqnarray}
S &=& \int \frac{d\omega d^{2}\mathbf{k}}{(2\pi)^3}
\Psi^\dagger_{\alpha}(i\omega,\mathbf{k})\left[i\omega-v_{F}(k_x\sigma_1
+ k_y\sigma_2)\right]\Psi_{\alpha}(i\omega,\mathbf{k}) \nonumber \\
&& + \frac{u\Lambda^2}{4}\int \frac{d\omega d\omega'
d^{2}\mathbf{k}d\Omega d^{2}\mathbf{q}}{(2\pi)^7}
\Psi^\dagger_{\alpha}(i\omega,\mathbf{k}\uparrow)
\sigma_2\Psi^\dagger_{\alpha}(\omega',-\mathbf{k}\downarrow)
\Psi_{\alpha}(\Omega,-\mathbf{q}\downarrow)\sigma_2
\Psi_{\alpha}(i\omega+i\omega'-i\Omega,\mathbf{q}\uparrow) \nonumber
\\
&& +\sum_{\Gamma}\frac{\Delta_{\Gamma}}{2} \int\frac{d\omega_1
d\omega_2d^{2}\mathbf{k}_1 d^{2}\mathbf{k}_2
d^{2}\mathbf{k}_3}{(2\pi)^8}
\Psi^\dagger_{\alpha}(i\omega_1,\mathbf{k}_1)\Gamma
\Psi_{\alpha}(i\omega_1,\mathbf{k}_2)
\Psi^\dagger_{\beta}(i\omega_2,\mathbf{k}_3)\Gamma
\Psi_{\beta}(i\omega_2,\mathbf{k}_1+\mathbf{k}_2+\mathbf{k}_3).
\end{eqnarray}
\end{widetext}
This action has been obtained by applying the replica trick to
average over random potential $A(\mathbf{r})$, with $\alpha$ and
$\beta$ being two replica indices and $\Delta_{\Gamma} = \Delta
v_{\Gamma}^2$. To distinguish different random potentials, we
introduce three new parameters $\Delta_M$, $\Delta_S$, and
$\Delta_V$ to characterize the effective strength of the
four-fermion couplings generated after averaging over random mass,
random chemical potential, and random gauge potential, respectively.
Notice that the coupling $u$ multiples a factor $\Lambda^2$, whose
meaning will be explained in Sec.~\ref{Sec_one_loop}.

The first term is the free fixed point of the action, and should be
kept invariant under the following re-scaling transformations
\begin{eqnarray}
k_{i} &=& k'_{i}e^{-l},\\
\omega &=& \omega'e^{-l},\\
\Psi_{\alpha}(i\omega,\mathbf{k}) &=&
\Psi_{\alpha}'(i\omega',\mathbf{k}')e^{2l},
\end{eqnarray}
where $l$ is a freely varying length scale. We will examine how the
other two interaction terms are modified under these transformations
in the next two sections.

\section{Cooper pairing in the clean limit}\label{Sec_clean}

We first consider the case of clean Dirac semimetals. The existence of a critical 
strength of attractive interaction, namely $u_c$, is well-known, and has been obtained 
previously by various methods~\cite{Uchoa07,CastroNeto,Sondhi13}. For completeness sake, 
we present the RG derivation of $u_c$ in this section and foresee the possible impact of disorders.

The leading correction to the fermion self-energy due to quartic
interaction is shown in Fig.~\ref{Fig_fermion_one_loop}. Using the
free fermion propagator
\begin{eqnarray}
G_0(i\omega,\mathbf{k}) = \frac{1}{i\omega -
v_{F}(k_x\sigma_1+k_y\sigma_2)},
\end{eqnarray}
it is easy to check that the self-energy is
\begin{eqnarray}
\Sigma_f\sim\mathrm{Tr}\int^{\infty}_{-\infty}
\frac{d\omega}{(2\pi)} \int_{b}^{1}
\frac{d^2\mathbf{k}}{(2\pi)^2}(-i\sigma_2)G_0(k) = 0.
\end{eqnarray}
This result simply implies that the quartic interaction does not
lead to renormalization of fermion velocity $v_F$, so we only need
to consider the renormalization of the coupling constant $u$.

\begin{figure}[htbp]
\center
\includegraphics[width=3.3in]{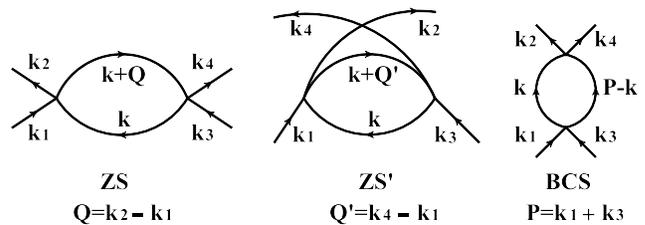}
\vspace{-0.2cm}\caption{One-loop vertex corrections to the pairing
interaction term, known as ZS, ZS', and BCS diagrams respectively.
The fourth momentum is determined according to momentum
conservation, namely $\mathbf{k}_4=\mathbf{k}_1 + \mathbf{k}_3 -
\mathbf{k}_2$.} \label{Fig_ZS_BCS_diagrams}
\end{figure}

We now proceed to compute the one-loop corrections to the quartic
coupling term. There are three sorts of diagrams for this vertex
corrections, as shown in Fig.~\ref{Fig_ZS_BCS_diagrams}. Borrowing
the terminology of Shankar \cite{Shankar1994RMP}, these three
diagrams are dubbed ZS, ZS', and BCS diagrams, respectively. We find
it convenient to first consider BCS diagram, which yields
\begin{eqnarray}
u^{\mathrm{BCS}}_{\mathrm{1L}} &=&
4\left(\frac{u\Lambda^2}{4}\right)^2 \mathrm{Tr}
\int^{\infty}_{-\infty}\frac{d\omega'}{(2\pi)}\int_{b}^{1}
\frac{d^2\mathbf{k'}}{(2\pi)^2}\nonumber \\
&&\times \sigma_2 G_0(k')\sigma_2 G_0(\mathbf{P}-k') \nonumber \\
&=&\left(\frac{1}{b^{-2}}\frac{u\Lambda^2}{4}\right)
\left(\frac{u\Lambda^2}{4}\right)\frac{l}{\pi v_{F}},
\label{Eq_u_BCS_one_loop}
\end{eqnarray}
with momentum $\mathbf{P}=0$ in the Cooper channel \cite{Sondhi13}.
We then move to compute the contributions of the ZS and ZS' diagrams
\cite{Shankar1994RMP}. It is straightforward to obtain
\begin{eqnarray}
u^{\mathrm{ZS}}_{\mathrm{1L}} &=& -\frac{u^2_0}{2}\mathrm{Tr}
\int^{\infty}_{-\infty}\frac{d\omega'}{(2\pi)}\int_{b}^{1}
\frac{d^2\mathbf{k'}}{(2\pi)^2}\nonumber \\
&&\times(\sigma_2)G_0(k')(\sigma_2)G_0(k'+\mathbf{Q}), \\
u^{\mathrm{ZS'}}_{\mathrm{1L}} &=& \frac{u^2_0}{2}\mathrm{Tr}
\int^{\infty}_{-\infty}\frac{d\omega'}{(2\pi)}\int_{b}^{1}
\frac{d^2\mathbf{k'}}{(2\pi)^2}\nonumber\\
&&\times(\sigma_2)G_0(k')(\sigma_2)G_0(k'+\mathbf{Q'}),
\end{eqnarray}
where $\mathbf{Q} = \mathbf{k}_2-\mathbf{k}_1$ and $\mathbf{Q'} =
\mathbf{k}_4-\mathbf{k}_1$, also defined in
Fig.~\ref{Fig_ZS_BCS_diagrams}. In ordinary metals which possess a
finite Fermi surface, the transferred momenta $Q$ and $Q'$ are
suppressed due to the large Fermi momentum, thus the ZS and ZS'
contributions are negligible compared to the BCS contribution
\cite{Shankar1994RMP}. In a Dirac fermion system, the Fermi momentum
$k_F \rightarrow 0$, and one needs to be more careful when dealing
with the ZS and ZS' diagrams. Since $\mathbf{k}_2$ and
$\mathbf{k}_1$ are both external momenta, they are much smaller than
the shell momenta to be integrated out in the process of carrying
out RG calculations \cite{Shankar1994RMP}. Accordingly, the
difference $\mathbf{Q} = \mathbf{k}_2 - \mathbf{k}_1$ can be
approximated as $\mathbf{Q} = 0$. Under these approximations, we
compute ZS diagram and get
\begin{eqnarray}
u^{\mathrm{ZS}}_{\mathrm{1L}} =
2u^{\mathrm{BCS}}_{\mathrm{1L}}.\label{Eq_ZS}
\end{eqnarray}
As for ZS' diagram, we assume a finite $Q'$ but henceforth utilize
$Q$ to substitute $Q'$ for notational simplicity. After introducing
a variable $\delta \equiv 2Q - Q^2$ with $Q\in(0,\sqrt{2}b)$ and
carrying straightforward calculations, we obtain
\begin{eqnarray}
u^{\mathrm{ZS'}}_{\mathrm{1L}}
&\equiv&-2u^{\mathrm{BCS}}_{\mathrm{1L}}f_Q,
\end{eqnarray}
where
\begin{eqnarray}
f_Q &\equiv& \frac{2-2(1-\delta)^{\frac{3}{2}}}{3\delta} +
\frac{4(Q^2_x-Q^2_y)}{15\delta^3}\nonumber \\
&&\times\left[4-5\delta+\sqrt{1-\delta}\left(\delta^2 + 3\delta -
4\right)\right].
\end{eqnarray}

Summing over the contributions from BCS, ZS, and ZS' diagrams yields
\begin{eqnarray}
u_{\mathrm{1L}} &=& u^{\mathrm{BCS}}_{\mathrm{1L}} +
u^{\mathrm{ZS}}_{\mathrm{1L}}
+ u^{\mathrm{ZS'}}_{\mathrm{1L}}\nonumber\\
&=& \left[1 + 2\left(1-f_Q\right)\right]
u^{\mathrm{BCS}}_{\mathrm{1L}}.
\end{eqnarray}
Since the system preserves translational symmetry, one can show that
\begin{eqnarray}
u_{\mathrm{1L}} = \left[1 + 2\left(1-f'_Q\right)\right]
u^{\mathrm{BCS}}_{\mathrm{1L}},
\end{eqnarray}
where
\begin{eqnarray}
u^{\mathrm{BCS}}_{\mathrm{1L}} &=& \frac{u^2_0}{16\pi v_{F}}l,
\\
f'_Q &\equiv& \frac{2 - 2(1-\delta)^{\frac{3}{2}}}{3\delta}.
\end{eqnarray}
Since the transferred momentum $Q$ is very small, it is easy to
verify that
\begin{eqnarray}
\lim_{Q \rightarrow 0}f'_Q = \lim_{\delta\rightarrow 0}f'_Q = 1,
\end{eqnarray}
which immediately indicates that
\begin{eqnarray}
\lim_{Q \rightarrow 0}u_{\mathrm{1L}} =
u^{\mathrm{BCS}}_{\mathrm{1L}}.
\end{eqnarray}

From the above calculations, we infer that the ZS and ZS'
contributions cancel each other provided that the external momenta
are sufficiently small. Since our focus is on the low-energy
asymptotic behaviors of the system, we will neglect the ZS and ZS'
diagrams in the next two sections and retain only the BCS diagram.
For completeness, we will revisit the effects of ZS and ZS' diagrams
in Sec.~\ref{Sec_ZS_ZS_prime}, where it will be showed that
including their contributions does not alter our basic conclusion.

\begin{figure}[htbp]
\center
\includegraphics[width=3.0in]{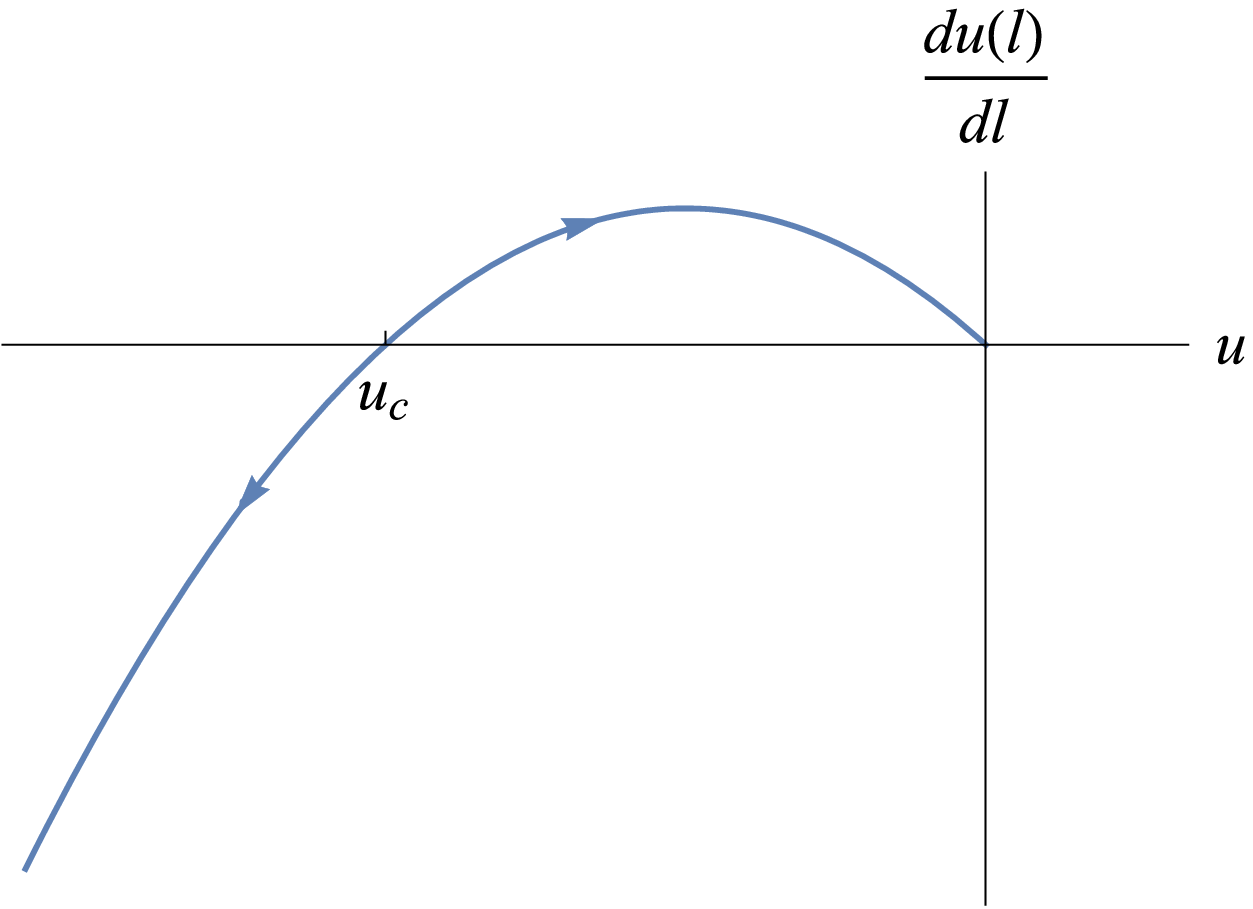}
\vspace{-0.20cm}\caption{Critical value $u_c$ is an unstable
infrared fixed point. Upon leaving this point, the BCS pairing
coupling either flows to zero, or grows monotonously, leading to
Cooper pairing instability.}\label{Fig_beta_u-u}
\end{figure}

Discarding the ZS and ZS' diagrams and adding the vertex correction
induced by the BCS diagram to the bare $u$ term, we find that the
BCS coupling flows according to the following equation:
\begin{eqnarray}
\frac{du}{dl} = -\left(1+\frac{u}{8\pi v_F}\right)u,
\end{eqnarray}
The critical coupling can be easily obtained from this equation:
\begin{eqnarray}
u_{c} = -8\pi v_F.
\end{eqnarray}
The corresponding flow diagram is presented in
Fig.~\ref{Fig_beta_u-u}. If the bare value $|u_0| < |u_{c}|$, the
pairing interaction flows to the trivial fixed point and Cooper
pairing cannot be formed. On the contrary, if $|u_0| > |u_{c}|$, the
attractive interaction flows to the strong coupling regime, which
leads to Cooper pairing instability.

The above results are not new and have already been obtained
previously by various approaches \cite{Uchoa07, Zhao06, Honerkamp08,
Sondhi13}. In the next section, we will include three types of
disorder and study their interplay with the pairing interaction by
carrying out detailed RG calculations. In that case, the flow
equation of $u$ might be substantially influenced by disorders, and,
as a consequence, superconductivity might be enhanced or suppressed.

\section{RG calculations in disordered Dirac semimetals}
\label{Sec_one_loop}

In this section, we study the interplay of Cooper pairing and
disorder by performing detailed RG analysis. The aforementioned
three types of disorders are supposed to coexist in the system. The
impact of each disorder can be separately examined by removing the
rest two.

Following Nandkishore \emph{et al.} \cite{Sondhi13}, we wish to
start our analysis directly from an effective BCS-type interaction
term that includes only the pairing between two Dirac fermions with
opposite momenta and spin (in case of singlet pairing). To this end,
we need to project the interaction term (\ref{Eq_H_int_eff}) onto
the Cooper channel \cite{Sondhi13}, which is justified because the
ZS and ZS' diagrams cancel each other at low energies. This can be
formally achieved by introducing a delta function
$\delta^2(\mathbf{p}+\mathbf{k})$ to $H_{\mathrm{int}}$ and then
integrate over $\mathbf{p}$. However, since a delta function
$\delta^2(\mathbf{p})$ scales like $\mathbf{p}^{-2}$, it might alter
the dimension of the coupling constant $u$. To solve this problem,
here we introduce an UV cutoff $\Lambda$ and write the effective BCS
interaction as
\begin{eqnarray}
H_{\mathrm{BCS}} = \frac{u \Lambda^2}{4}
\sum_{\mathbf{k},\mathbf{q}}
\Psi^\dagger_{\mathbf{k},\uparrow}(-i\sigma_2)
\Psi^\dagger_{-\mathbf{k},\downarrow}
\Psi_{-\mathbf{q},\downarrow}(i\sigma_2)
\Psi_{\mathbf{q},\uparrow}.\label{Eq_H_BCS}
\end{eqnarray}
Here, $\Lambda$ can be considered as the contributions from the
neglected non-BCS coupling terms. It should scale as $\mathbf{p}^2$
and becomes progressively unimportant as one goes to lower and lower
energies. An alternative approach is to regard Eq.~(\ref{Eq_H_BCS})
as the starting point and define a new effective coupling constant
$u_{\mathrm{eff}} = u \Lambda^2/4$, which will lead us to the same
results.

It is necessary to pause here and briefly remark on the validity of
introducing the above attractive interaction term. To acquire a net
attraction between Dirac fermions, the attractive force mediated by
either phonons or plasmons should be larger than the Coulombic
repulsive force \cite{Uchoa07}. Due to the vanishing of zero-energy
DOS, the Coulomb interaction is only poorly screened by the
particle-hole continuum \cite{CastroNeto, Kotov} and thus makes it
hard to achieve a net attraction. However, the strength of Coulomb
interaction can be substantially reduced when the Dirac semimetal is
placed on some metallic substrate \cite{CastroNeto, Kotov,
DasSarma}. Moreover, disorders may generate a finite DOS at the
Dirac points, which also strongly suppresses the Coulomb interaction
via static screening \cite{Liu09, LiuNJP11}. Therefore, it is in
principle possible for Dirac semimetals to develop a net attractive
interaction. Our following analysis will be based on the assumption
that a net attraction is realized in an intrinsic 2D Dirac
semimetal.

\begin{figure}[htbp]
\center
\includegraphics[width=2.5in]{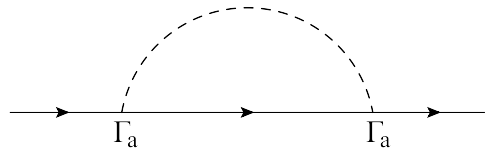}
\vspace{-0.05cm}\caption{One-loop correction to fermion propagator
due to disorder scattering. Here, the dashed line represents the
disorder scattering and $\Gamma_a$ should sum over all the three
types of disorder.}\label{Fig_fermion_dis_one_loop}
\end{figure}

\begin{figure*}[htbp]
\center
\includegraphics[width=6.6in]{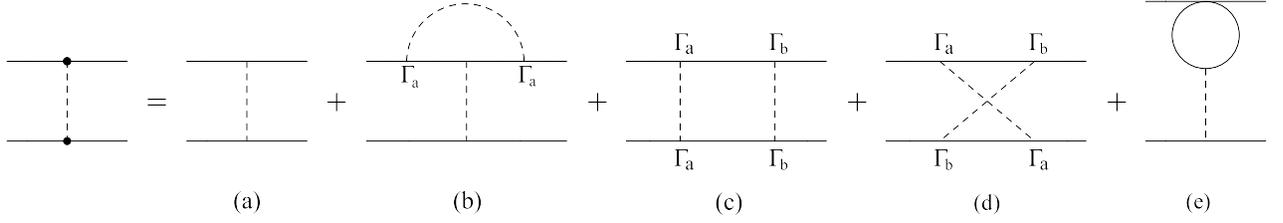}
\caption{One-loop corrections to the disorder strength in the
replica limit. The repeated $\Gamma_{a,b}$ should sum over all the
three types of disorder.}\label{Fig_fermion_dis_vertex_one_loop}
\end{figure*}
\begin{figure}[htbp]
\center
\includegraphics[width=3.2in]{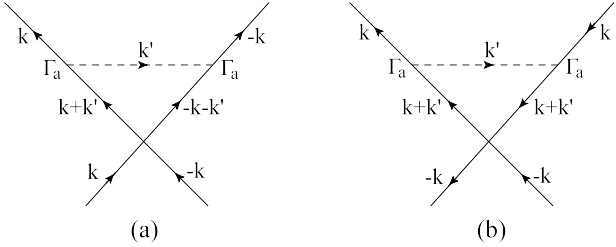}
\vspace{-0.39cm}\caption{One-loop vertex corrections to the pairing
interaction due to disorder scattering.}\label{Fig_u0_dis_one_loop}
\end{figure}

In the presence of disorders, the Dirac fermions receive additional
self-energy corrections due to disorder scattering. The leading
correction presented in Fig.~\ref{Fig_fermion_dis_one_loop} leads to
\begin{eqnarray}
\Sigma_{\mathrm{dis}}(i\omega) &=&\sum_{\Gamma} \Delta_{\Gamma}
\int\frac{d^2\mathbf{q}}{(2\pi)^2}\Gamma
G_{0}(i\omega,\mathbf{q})\Gamma \nonumber \\
&=& -i\omega \frac{\Delta_M + \Delta_S + 2\Delta_V}{2\pi v^2_{F}} l.
\label{Eq.Self_energy_d}
\end{eqnarray}

To study the renormalization of disorder parameter
$\Delta_{\Gamma}$, we next would consider the vertex corrections to
the effective quartic coupling induced by disorder averaging
procedure, as schematically shown in
Fig.~\ref{Fig_fermion_dis_vertex_one_loop}. Clearly,
Fig.~\ref{Fig_fermion_dis_vertex_one_loop}(a) represents the bare
vertex, and we only need to compute the rest four diagrams given by
Fig.~\ref{Fig_fermion_dis_vertex_one_loop}(b-e). The contribution of
Fig.~\ref{Fig_fermion_dis_vertex_one_loop}(b) is given by
\begin{eqnarray}
\delta\Delta_{\Gamma}^{b}&=& \Delta_{\Gamma} \sum_{\Gamma_b}\Delta_b
\int\frac{d^2\mathbf{q}}{(2\pi)^2}\Gamma_b G_0(\mathbf{q}) \Gamma
G_0(\mathbf{q})\Gamma_b.
\end{eqnarray}
The matrix $\Gamma$ has different expression in the case of
different types of disorder. For random chemical potential,
$\Gamma=\sigma_0$ and we have
\begin{eqnarray}
\delta\Delta_{S}^{b} = \frac{(\Delta_S+\Delta_{M}+
2\Delta_V)\Delta_S}{2\pi v_{F}^{2}}\sigma_0 l.
\label{Eq.Disorder_Vertex_b_chem}
\end{eqnarray}
For random mass, $\Gamma=\sigma_3$ and we have
\begin{eqnarray}
\delta\Delta_{M}^{b} &=& \frac{-(\Delta_S+\Delta_{M} -
2\Delta_V)\Delta_{M}}{2\pi v_{F}^{2}}\sigma_3 l.
\label{Eq.Disorder_Vertex_b_mass}
\end{eqnarray}
For random gauge potential, there are two components, namely
$\Gamma=\sigma_1$ and $\Gamma=\sigma_2$. In both of these two cases,
we find that
\begin{eqnarray}
\delta\Delta_{V}^{b} = 0.
\end{eqnarray}
The contributions form Fig.~\ref{Fig_fermion_dis_vertex_one_loop}(c)
and Fig.~\ref{Fig_fermion_dis_vertex_one_loop}(d) are best computed
at once. It is convenient to sum them up and obtain
\begin{eqnarray}
\delta\Delta_{\Gamma}^{c+d} &=& \sum_{\Delta_a}\sum_{\Delta_b}
\Delta_a \Delta_b \int\frac{d^2\mathbf{p}}{(2\pi)^2}
\psi^\dagger_{\alpha}[\Gamma_a G_0(0,\mathbf{p})\Gamma_b]\psi_{\alpha}
\nonumber \\
&& \times \psi^\dagger_{\beta}[\Gamma_b G_0(0,\mathbf{p})\Gamma_a +
\Gamma_a G_0(0,-\mathbf{p})\Gamma_b]\psi_{\beta}.
\label{Eq.Disorder_Vertex_c_0}
\end{eqnarray}
Straightforward calculations yield
\begin{eqnarray}
\delta\Delta_{S}^{c+d} &=& \frac{2\Delta_M \Delta_V}{2\pi v_{F}^2}l
(\bar{\psi}_{\alpha}\sigma_0\psi_{\alpha})(\bar{\psi}_{\beta}\sigma_0\psi_{\beta}),
\label{Eq.Disorder_Vertex_d_1}\\
\delta\Delta_{M}^{c+d} &=& \frac{2\Delta_S \Delta_V}{2\pi v_{F}^2}l
(\bar{\psi}_{\alpha}\sigma_3\psi_{\alpha})(\bar{\psi}_{\beta}\sigma_3\psi_{\beta}),
\label{Eq.Disorder_Vertex_d_2}\\
\delta\Delta_{V}^{c+d} &=& \frac{\Delta_M \Delta_S}{2\pi v_{F}^2}l
(\bar{\psi}_{\alpha}\sigma_j\psi_{\alpha})(\bar{\psi}_{\beta}\sigma_j\psi_{\beta}),
\label{Eq.Disorder_Vertex_d_3}
\end{eqnarray}
which apply to the case of random chemical potential, random mass,
and random gauge potential, respectively. Here, the repeated index
$j$ sums over the two components of random gauge potential. For all
the other cases with $\Gamma_a = \Gamma_b = \sigma_{0,1,2,3}$, these
two diagrams cancel each other and make no contributions to the
vertex.

There is now only one diagram left, given by
Fig.~\ref{Fig_fermion_dis_vertex_one_loop}(e). Similar to the
one-loop correction to the coupling $u$, there are three
possibilities, corresponding to ZS, ZS', and BCS like diagrams, as
explicitly shown in Ref.~\cite{Sondhi13}. As we have illustrated in
Sec.~\ref{Sec_clean}, the ZS and ZS' diagrams cancel each other and
the BCS diagram makes no contribution because the loop momentum lies
in the slim shell due to the momentum restriction, as showed by
Fig.\ref{Fig_fermion_one_loop}. Finally, as argued in
Ref.~\cite{Sondhi13}, this sort of diagram simply vanishes and
contributes nothing to the quartic coupling term represented by
parameter $\Delta_{\Gamma}$.

Apart from the fermion-disorder vertex corrections, there are two
one-loop diagrams contributing to the BCS interaction $u$ due to
disorders, as given by Fig.~\ref{Fig_u0_dis_one_loop}. It is easy to
find that
\begin{eqnarray}
u^{\mathrm{1L}}_{\mathrm{dis1}} &=& \frac{u}{4}\left(
\frac{\Delta_M+\Delta_S+2\Delta_V}{4\pi v_{F}}\right), \\
u^{\mathrm{1L}}_{\mathrm{dis2}} &=& -\frac{u}{4}\left(
\frac{\Delta_M+\Delta_S+2\Delta_V}{4\pi v_{F}}\right).
\end{eqnarray}
Apparently, these two contributions cancel precisely, and thus can
be simply dropped.

Now we have evaluated all the leading corrections to fermion
self-energy and disorder vertex, and are ready to derive the RG
equations. To proceed, we need to integrate out the modes defined in
the momentum shell $b\Lambda < k < \Lambda$, where $b$ can be
written as $b = e^{-l}$. Under the scaling transformation $k_{i} =
k'_{i}e^{-l}$ and $\omega = \omega'e^{-l}$, the fermion field and
disorder potential should transform as follows \cite{Huh2008}
\begin{eqnarray}
\Psi_{\alpha}(i\omega,\mathbf{k}) &=&
\Psi_{\alpha}'(i\omega',\mathbf{k}')
e^{\frac{1}{2}\int_{0}^{l}dl(4-\eta)}.
\end{eqnarray}
where $\eta$ is an anomalous dimension for the fermion field $\Psi$
induced by disorders.

To compute $\eta$, we first redefine the effective parameter for
random potential as follows
\begin{eqnarray}
\frac{\Delta_{\Gamma}}{2\pi v_{F}^2} \rightarrow
\Delta_{\Gamma}.\label{Eq.dis_coup}
\end{eqnarray}
Adding the fermion self-energy $\Sigma_{\mathrm{dis}}(i\omega) =
-i\omega (\Delta_M+\Delta_S+2\Delta_V) l$ to the free fermion
action, we have
\begin{eqnarray}
&&\int^b_0\frac{d\omega}{2\pi}\frac{d^{2}\mathbf{k}}{(2\pi)^2}
\Psi^\dagger_{\alpha}[1+(\Delta_M+\Delta_S+2\Delta_V) l](i\omega)
\Psi_{\alpha}\nonumber \\
&=&\int^b_0\frac{d\omega}{2\pi}\frac{d^{2}\mathbf{k}}{(2\pi)^2}
\Psi^\dagger_{\alpha}(i\omega)e^{(\Delta_M+\Delta_S+2\Delta_V)l}
\Psi_{\alpha},
\end{eqnarray}
which after rescaling transformations becomes
\begin{eqnarray}
\int^1_0\frac{d\omega}{2\pi}\frac{d^{2}\mathbf{k}}{(2\pi)^2}
\Psi^\dagger_{\alpha}(i\omega)e^{-\eta l +
(\Delta_M+\Delta_S+2\Delta_V)l}\Psi_{\alpha}.
\end{eqnarray}
This term is required to return to its original form, which forces
us to demand that
\begin{eqnarray}
\eta = \Delta_M+\Delta_S+2\Delta_V.\label{Eq_eta}
\end{eqnarray}

By using the above anomalous dimension and the one-loop quantum
corrections we have just computed, we eventually obtain the
following RG equations:
\begin{eqnarray}
\frac{dv_F}{dl} &=& \!\!-(\Delta_M + \Delta_S + 2\Delta_V)v_F,
\label{Eq:MultiRGVF} \\
\frac{d\Delta_S}{dl} &=& \!\!2(\Delta_S + 2\Delta_V + \Delta_M)\Delta_S
+ 4\Delta_M\Delta_V,\label{Eq:MultiRGVGamma1} \\
\frac{d\Delta_M}{dl} &=& \!\!-2(\Delta_M - 2\Delta_V + \Delta_S)\Delta_M
+ 4\Delta_S\Delta_V,
\label{Eq:MultiRGVGamma2}\\
\frac{d\Delta_V}{dl} &=& \!\!2\Delta_S\Delta_M,
\label{Eq:MultiRGVGamma3} \\
\frac{du}{dl} &=& \!\!-\left[1+2(\Delta_M + \Delta_S + 2\Delta_V) +
\frac{u}{8\pi v_F}\right]\!\!u. \label{Eq:MultiRGu}
\end{eqnarray}
We notice that
Eqs.~(\ref{Eq:MultiRGVGamma1})-(\ref{Eq:MultiRGVGamma3}) are in
accordance with the results obtained previously in Refs.
\cite{Aleiner2006PRL, Aleiner2008PRB, Foster12}. In the next
section, we will use these RG equations to analyze how disorders
affect the formation of superconductivity.

\section{RG analysis of the disorder effects on superconductivity}\label{Sec_one_disorder}

In this section, we will first consider the impact of each single
disorder on superconductivity by simply removing the rest two types
of disorder from the complete set of RG equations. We pay special
attention to the (ir)relevance of the effective disorder parameter
$\Delta_{\Gamma}$ and how the BCS coupling $u$ is modified by the
disorder. In addition, we are also interested in the low-energy
behaviors of some physical quantities, including the fermion
velocity $v_{F}$, quasi-particle residue $Z_f$, and fermion DOS
$\rho(\omega)$.

\begin{figure*}[htbp]
\includegraphics[width=2.3in]{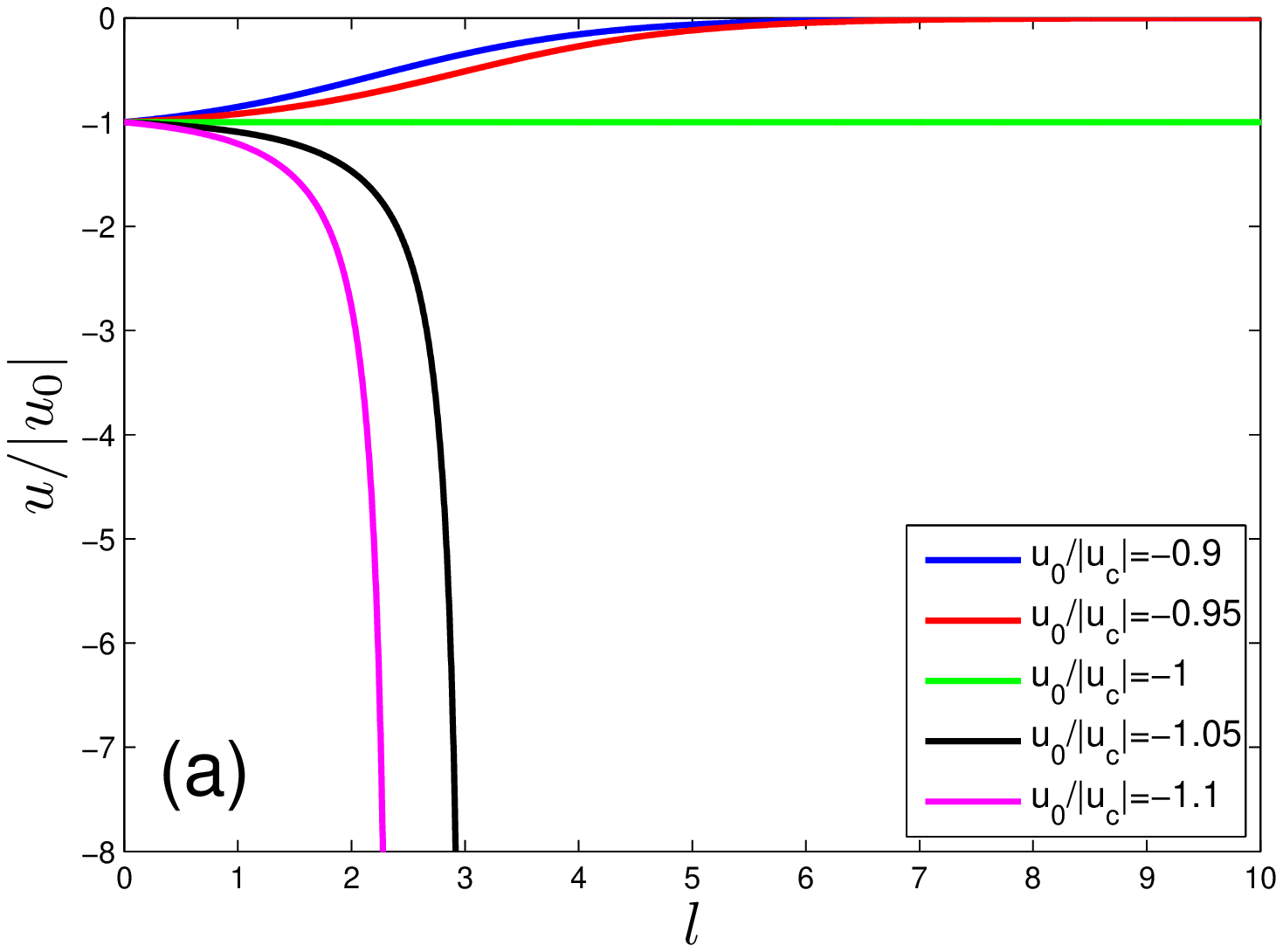}
\includegraphics[width=2.3in]{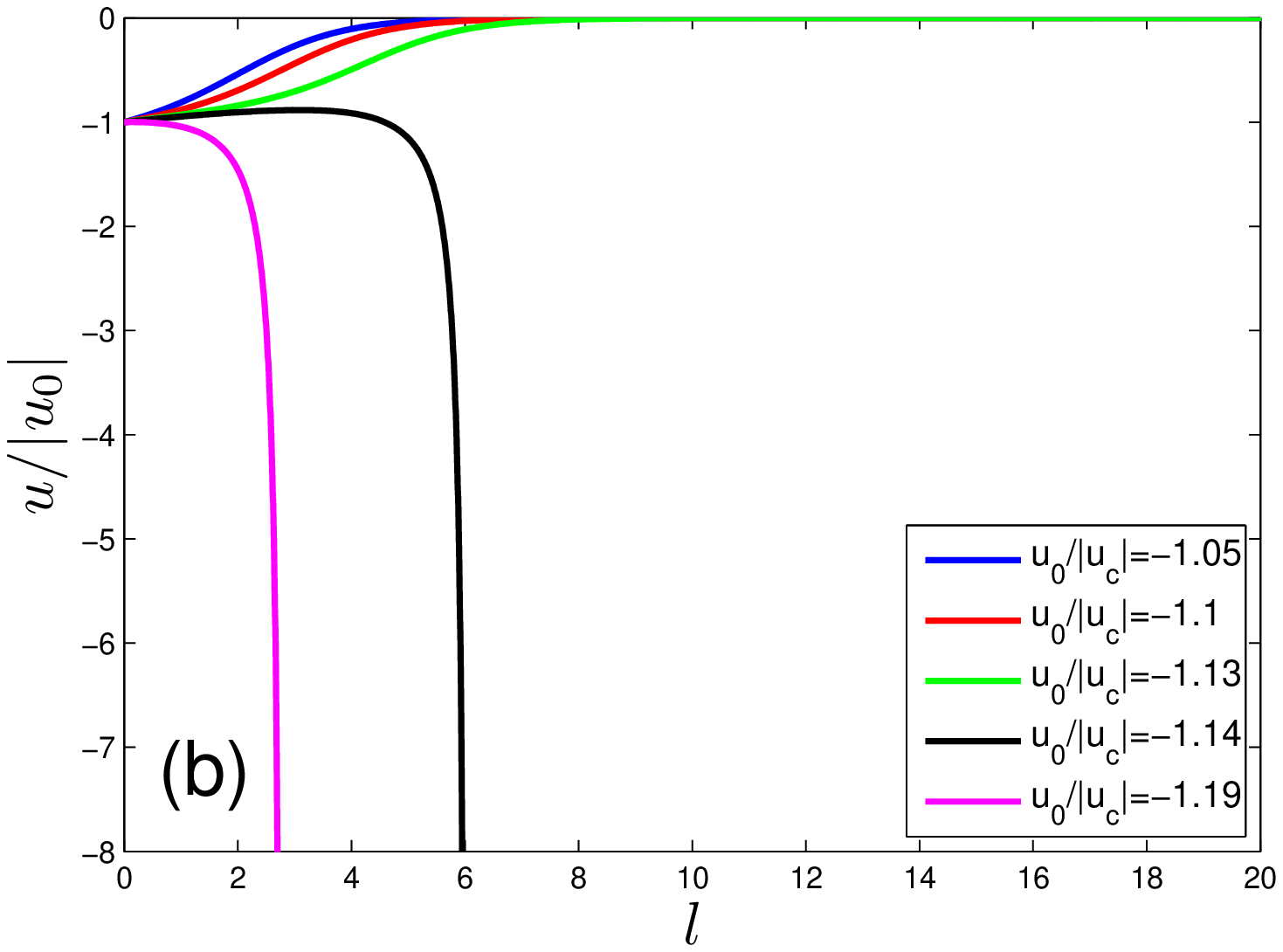}
\includegraphics[width=2.3in]{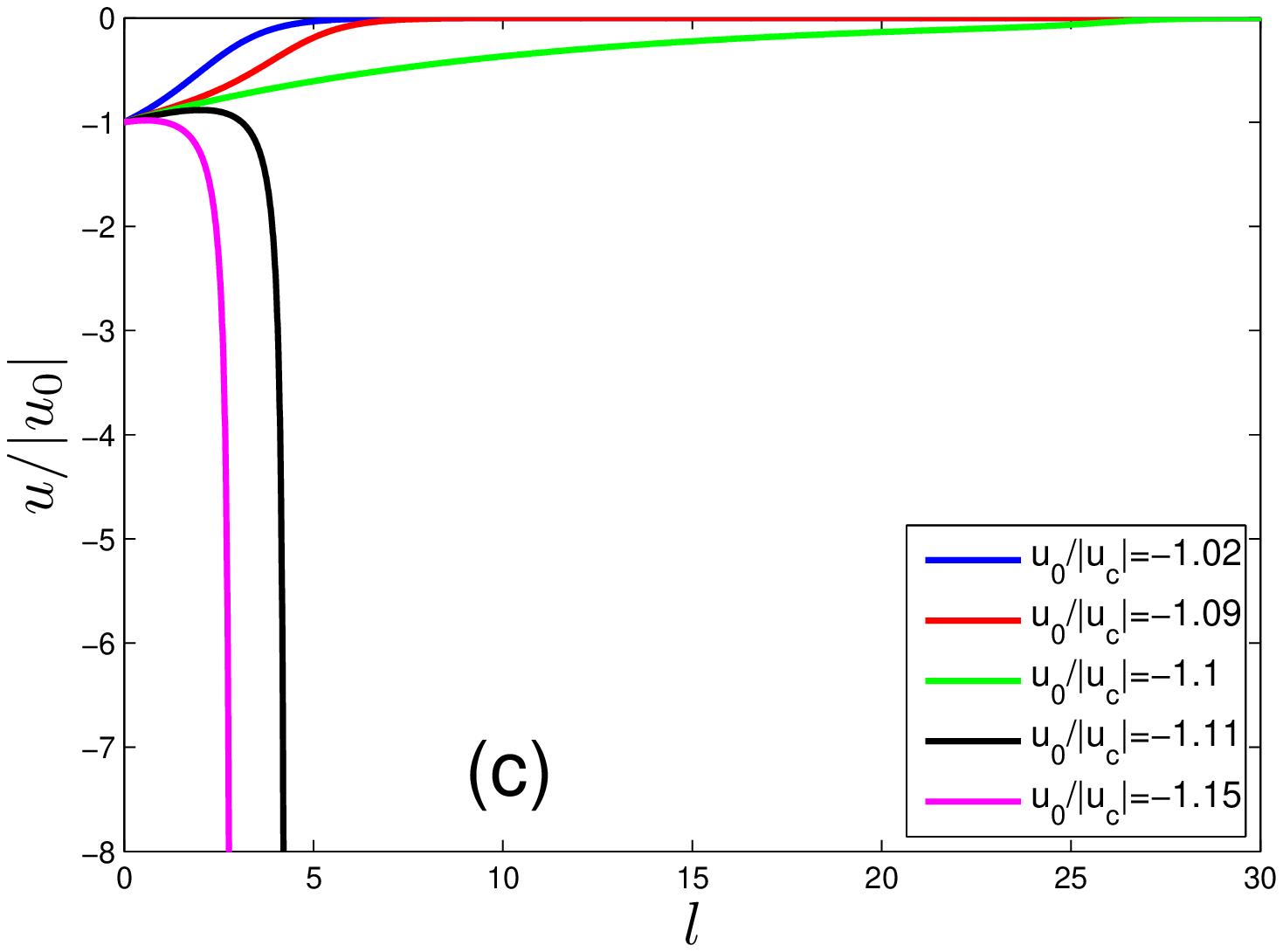}
\caption{(Color online) Flow of pairing interaction parameter $u$ in
the case of: (a) clean limit; (b) random mass with $\Delta_{M}^0 =
0.1$; (c) random gauge potential with $\Delta_{V}^0 = 0.05$. The
parameter $u$ with a small initial value flows to zero at large $l$,
whereas a sufficiently large $u$ flows rapidly to strong coupling,
signalling the onset of superconductivity. A critical value for $u$
exists in each case, but the absolute value $|u_c'|$ obtained in the
presence of disorder is always larger than $|u_c|$ obtained in clean
limit. It can be verified that stronger disorder gives rise to
larger value of $|u_c'|$, which further suppresses
superconductivity.} \label{Fig:vRGuFlow}
\end{figure*}

\subsection{Random mass}\label{Sec_RM}

In the case that random mass exists alone in the system, we can
simply set $\Delta_S = \Delta_V = 0$. Now, the complete set of RG
equations are simplified to
\begin{eqnarray}
\frac{dv_F}{dl} &=& -\Delta_M v_F,\label{Eq:MultiRGVF_Sim} \\
\frac{d\Delta_M}{dl} &=&-2\Delta_M^2, \label{Eq:MultiRGVGamma2_Sim}\\
\frac{du}{dl} &=& -\left[1+2\Delta_M+\frac{u}{8\pi v_F}\right]u.
\label{Eq:MultiRGu_sim}
\end{eqnarray}
It is clear that the velocity $v_F$ and disorder parameter
$\Delta_{M}$ are no longer constants, but flow with varying length
scale $l$ according to Eq.~(\ref{Eq:MultiRGVF_Sim}) and
Eq.~(\ref{Eq:MultiRGVGamma2_Sim}). To determine the new critical
value $|u_{c}'|$, we need to solve these two RG equations. It is
easy to find that Eq.~(\ref{Eq:MultiRGVGamma2_Sim}) has a solution
\begin{eqnarray}
\Delta_M(l) =
\frac{\Delta_{M}^0}{1 + 2\Delta_{M}^0l}.\label{Eq:CgMassFlow}
\end{eqnarray}
Substituting Eq.~(\ref{Eq:CgMassFlow}) to
Eq.~(\ref{Eq:MultiRGVF_Sim}) and solving the differential equation
gives rise to
\begin{eqnarray}
v_{F}(l) = \frac{v_{F}^0}{\sqrt{2\Delta_{M}^0l + 1}}.
\end{eqnarray}
Apparently, $\Delta_{M}$ and $v_F$ decrease very slowly with growing
$l$ and eventually vanish as $l \rightarrow +\infty$. After
including random mass, the flow of $u$ for different values of
$u_{0}/|u_{c}|$ at $\Delta_{M}^0=0.1$ is shown in
Fig.~\ref{Fig:vRGuFlow}(b). In the clean limit, for the specific
initial values of $u_{0}/|u_{c}| = -1.05$, $u_{0}/|u_{c}| = -1.1$,
and $u_{0}/|u_{c}| = -1.13$, $u$ flows rapidly to the strong
coupling regime, which signals the onset of superconductivity.
However, after random mass is included, $u$ flows to zero in the
lowest-energy limit starting from the same initial values, which
implies that the system remains gapless. It is therefore clear that
random mass tends to suppress superconductivity. For larger initial
values of $u_{0}/|u_{c}|$, $u$ still flows to strong coupling. The
new QCP is located at $u_{c}'$, which is greater than $u_c$.

The above RG analysis show that the effective disorder parameter
$\Delta_{M}$ vanishes ultimately as $l \rightarrow +\infty$.
However, $\Delta_{M}$ flows to zero slowly with growing $l$.
Concretely, according to Eq.~(\ref{Eq:CgMassFlow}), we find that
$\Delta_{M} \sim \frac{1}{l}$. In the spirit of RG theory, this
means that random mass is marginally irrelevant in a 2D Dirac
fermion system. Nevertheless, before $\Delta_{M}$ flowing to zero,
random mass can still induce weak corrections to observable
quantities of the system. As shown in the above analysis, random
mass drives $v_{F}$ to vanish at very low energies and increases the
critical BCS coupling $u_c$.

We now analyze three important quantities, namely the Landau damping
rate, the quasiparticle residue $Z_{f}$, and the low-energy DOS. The
residue is usually defined as
\begin{eqnarray}
Z_f = \frac{1}{1 - \frac{\partial
\mathrm{Re}\Sigma^{R}(\omega)}{\partial\omega}}\label{Eq.def_Z},
\end{eqnarray}
where $\mathrm{Re}\Sigma^{R}$ is the real part of retarded fermion
self-energy. By virtue of the RG results and also according to the
one-loop self-energy given by Eq.~(\ref{Eq.Self_energy_d}), it is
convenient to express the residue in the following form
\cite{WangLiu2014,WangLiuZhang16}
\begin{eqnarray}
\frac{dZ_{f}}{dl} = -\Delta_M Z_{f}.\label{Eq:EqDfZf1}
\end{eqnarray}
Making use of Eq.~(\ref{Eq:CgMassFlow}), it is easy to find that
$Z_{f} \sim \frac{1}{\sqrt{2\Delta_{M}^{0}l + 1}}\rightarrow 0$ in
the limit $l \rightarrow +\infty$. Hence, the Dirac fermions are not
well-defined Landau quasiparticles. Using the scaling relation
$\omega = \omega_{0}e^{-l}$, where $\omega_0$ is a UV cutoff, we get
\begin{eqnarray}
\mathrm{Re}\Sigma^{R}(\omega) \sim \omega
\left[\ln\left(\frac{\omega_{0}}{\omega}\right)\right]^{\frac{1}{2}}.
\end{eqnarray}
According to the Kramers-Kronig (KK) relation, the imaginary part of
retarded self-energy depends on $\omega$ as
\begin{eqnarray}
\mathrm{Im}\Sigma^{R}(\omega) \sim
\frac{\omega}{\left[\ln\left(\frac{\omega_{0}}{\omega}\right)\right]^{\frac{1}{2}}},
\label{Eq:DampingRM}
\end{eqnarray}
which apparently is a MFL-like behavior. The RG equation for the
low-energy DOS is give by \cite{WangLiu2014, WangLiuZhang16}
\begin{equation}
\frac{d\ln\rho(\omega)}{d\ln(\omega)} = \frac{1 - \Delta_M}{1+\Delta_M}.
\label{Eq:RGRhoDef}
\end{equation}
After solving this equation, we obtain
\begin{eqnarray}
\rho(\omega) \sim \omega\ln\left(\frac{\omega_{0}}{\omega}\right).\label{Eq:DOSRM}
\end{eqnarray}
Comparing to the low-energy DOS $\rho(\omega)\sim \omega$ for clean,
non-interacting 2D Dirac semimetal, we find that the low-energy DOS
$\rho(\omega)$ is enhanced by random mass.

Based on the above analysis, we plot the schematic phase diagram
spanned by $\Delta_{M}^0$ and $u_0$ in
Fig.~\ref{Fig:PhaseDiagram}(a). In the clean limit, the critical
coupling $u_c$ defines a QCP between a non-interacting Dirac
semimetallic phase and a superconducting phase. In contrast, in the
presence of random mass, the new critical coupling $u_{c}'$, whose
absolute value is larger than $|u_{c}|$, corresponds to a QCP
between a MFL-like phase and a superconducting phase.

\subsection{Random gauge potential}\label{Sec_RVP}

Setting $\Delta_S = \Delta_M = 0$, the RG equations become
\begin{eqnarray}
\frac{dv_F}{dl} &=& -2\Delta_V v_F,\label{Eq:MultiRGVF_SimG} \\
\frac{d\Delta_V}{dl} &=&0, \label{Eq:MultiRGVGamma2_SimG}\\
\frac{du}{dl} &=& -\left[1+4\Delta_V+\frac{u}{8\pi v_F}\right]u.
\label{Eq:MultiRGu_simG}
\end{eqnarray}
We know from Eq.~(\ref{Eq:MultiRGVGamma2_SimG}) that random gauge
potential is marginal and $\Delta_V$ should be a constant, namely
\begin{eqnarray}
\Delta_V (l)= \Delta_V^0.\label{Eq:CgFlowGP}
\end{eqnarray}
Substitute this constant to Eq.~(\ref{Eq:MultiRGVF_SimG}), we get
\begin{eqnarray}
v_{F}(l) = v_{F}^0e^{-2\Delta_V^0l}.
\end{eqnarray}
Random gauge potential drives the fermion velocity to decay
exponentially, which in turn alters the critical coupling $u_c$. At
the chosen value $\Delta_V^0 = 0.1$, the flow of $u$ at different
initial values of $u_{0}/|u_{c}|$ is depicted in
Fig.~\ref{Fig:vRGuFlow}(c). We observe that random gauge potential
modifies the critical value $u_{c}$ to $u_{c}'$ with a larger
absolute value, and thus suppresses superconductivity. Based on
Eqs.~(\ref{Eq.def_Z}) and (\ref{Eq:CgFlowGP}), we find that the
residue behaves as
\begin{eqnarray}
Z_{f}(l) = e^{-2\Delta_V^0 l},
\end{eqnarray}
which vanishes rapidly with growing $l$. It is easy to obtain
\begin{eqnarray}
\mathrm{Im}\Sigma^{R}(\omega) \sim \omega^{1-2\Delta_V^0}. \label{Eq:DampingGP}
\end{eqnarray}
This is typical NFL behavior since $\Delta_V^0 > 0$. Using
Eqs.~(\ref{Eq:RGRhoDef}) and (\ref{Eq:CgFlowGP}), we get the
low-energy DOS
\begin{eqnarray}
\rho(\omega) = \omega^{\frac{1 - 2\Delta_V^0}{1 + 2\Delta_V^0}},\label{Eq:DOSGP}
\end{eqnarray}
which is enhanced by random gauge potential.

\begin{figure}[htbp]
\includegraphics[width=2.8in]{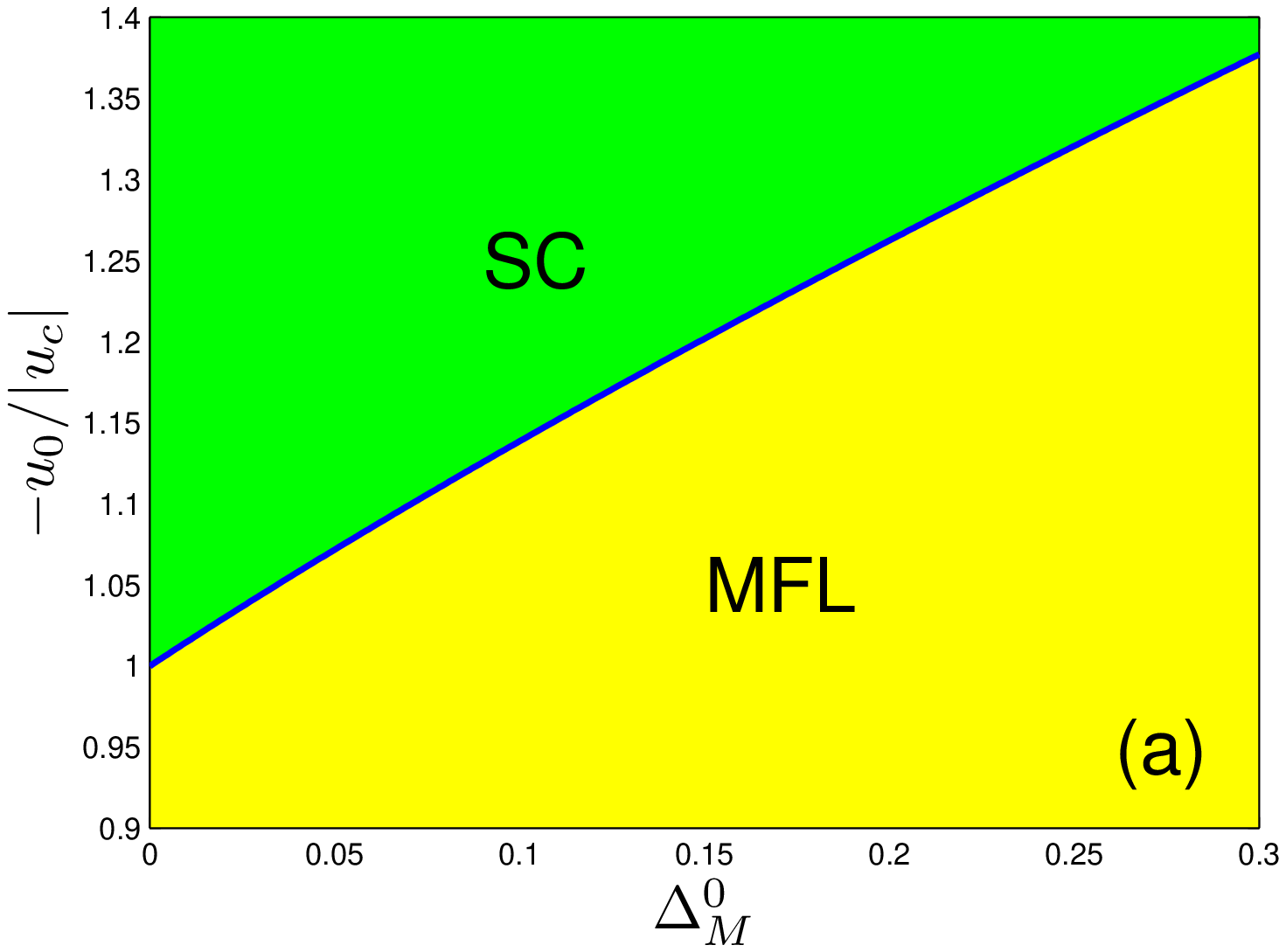}
\includegraphics[width=2.8in]{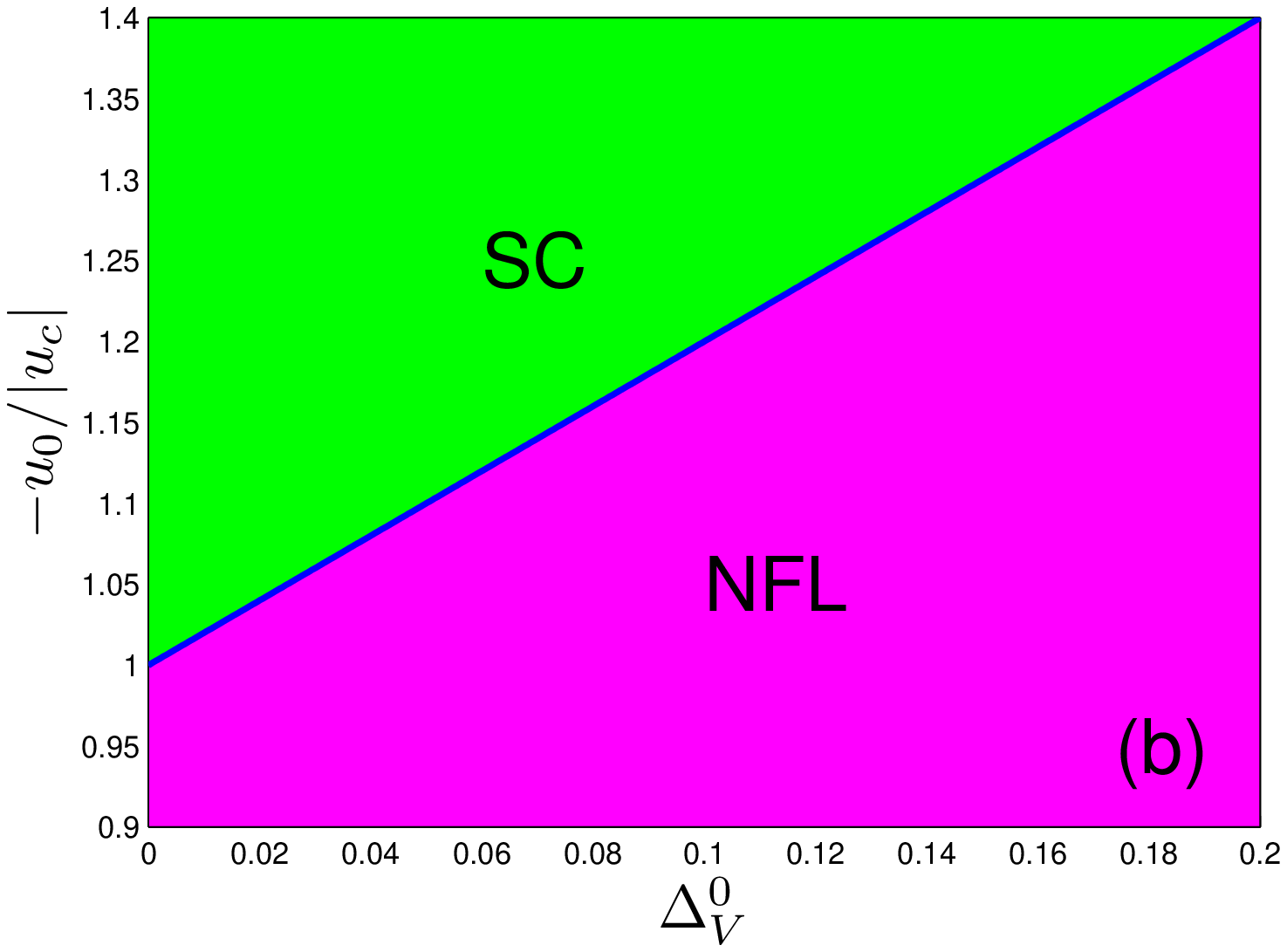}
\caption{(Color online) Phase diagram in the space spanned by
$\Delta_{\Gamma}$ and $u_{0}$ in the case of (a) random mass; (b)
random gauge potential. Here, SC refers to the superconducting
phase.} \label{Fig:PhaseDiagram}
\end{figure}

If we fix $\Delta_V^0$ and tune the coupling $u$, the system
undergoes a quantum phase transition between a NFL and a
superconducting phase, with $u_{c}'$ being the QCP. The schematic
phase diagram in the space spanned by $\Delta_V^0$ and $u_{0}$ is
shown in Fig.~\ref{Fig:PhaseDiagram}(b). There is a critical line on
the phase diagram, separating the superconducting phase from the NFL
phase.

\subsection{Random chemical potential}\label{Sec_RSP}

In the case of random chemical potential, the RG equations are
\begin{eqnarray}
\frac{dv_F}{dl} &=& -\Delta_S v_F,\label{Eq:MultiRGVF_SimS} \\
\frac{d\Delta_S}{dl} &=&2\Delta_S^2, \label{Eq:MultiRGVGamma2_SimS}\\
\frac{du}{dl} &=& -\left[1+2\Delta_S+\frac{u}{8\pi v_F}\right]u.
\label{Eq:MultiRGu_simS}
\end{eqnarray}
Similarly, solving Eq.~(\ref{Eq:MultiRGVGamma2_SimS}) gives
\begin{eqnarray}
\Delta_S(l) =\frac{\Delta_{S}^0}{1 - 2\Delta_{S}^0l}.\label{Eq:CgMassFlowS}
\end{eqnarray}
Substituting Eq.~(\ref{Eq:CgMassFlowS}) to
Eq.~(\ref{Eq:MultiRGVF_SimS}), and solving the differential equation
we get
\begin{eqnarray}
v_{F}(l) = v_{F}^0\sqrt{1-2\Delta_{S}^0l}.
\end{eqnarray}
There exists a characteristic length scale $l_{c} =
1/2\Delta_{s}^{0}$. As $l$ approaches $l_c$ from below, the
effective strength parameter $\Delta_{S}\rightarrow \infty$ and the
fermion velocity $v_{F}\rightarrow 0$. It is thus clear that random
chemical potential is a relevant perturbation to the system. This
behavior is usually interpreted as a signature that the Dirac
fermion system enters into a disorder-controlled diffusive state
\cite{Altland02}. However, since $\Delta_{S}$ flows to the strong
coupling, the perturbative RG method progressively breaks down and
cannot provide a clear answer to the fate of superconductivity.

Let us briefly summarize the RG results here. In the cases of random
mass and random gauge potential, the strength parameter
$\Delta_{\Gamma}$ either vanishes or can be fixed at certain small
value. Therefore, the conclusions that superconductivity is
suppressed and that the value of increased critical value $u_{c}'$
obtained by RG analysis are expected to be reliable. In the special
case of random chemical potential, however, the impact of random
chemical potential on superconductivity remains elusive. A more
efficient approach should be developed to address this issue, which
will be discussed in more detail in Sec.~\ref{Sec_discuss}.

\begin{figure*}[htbp]
\includegraphics[width=2.6in]{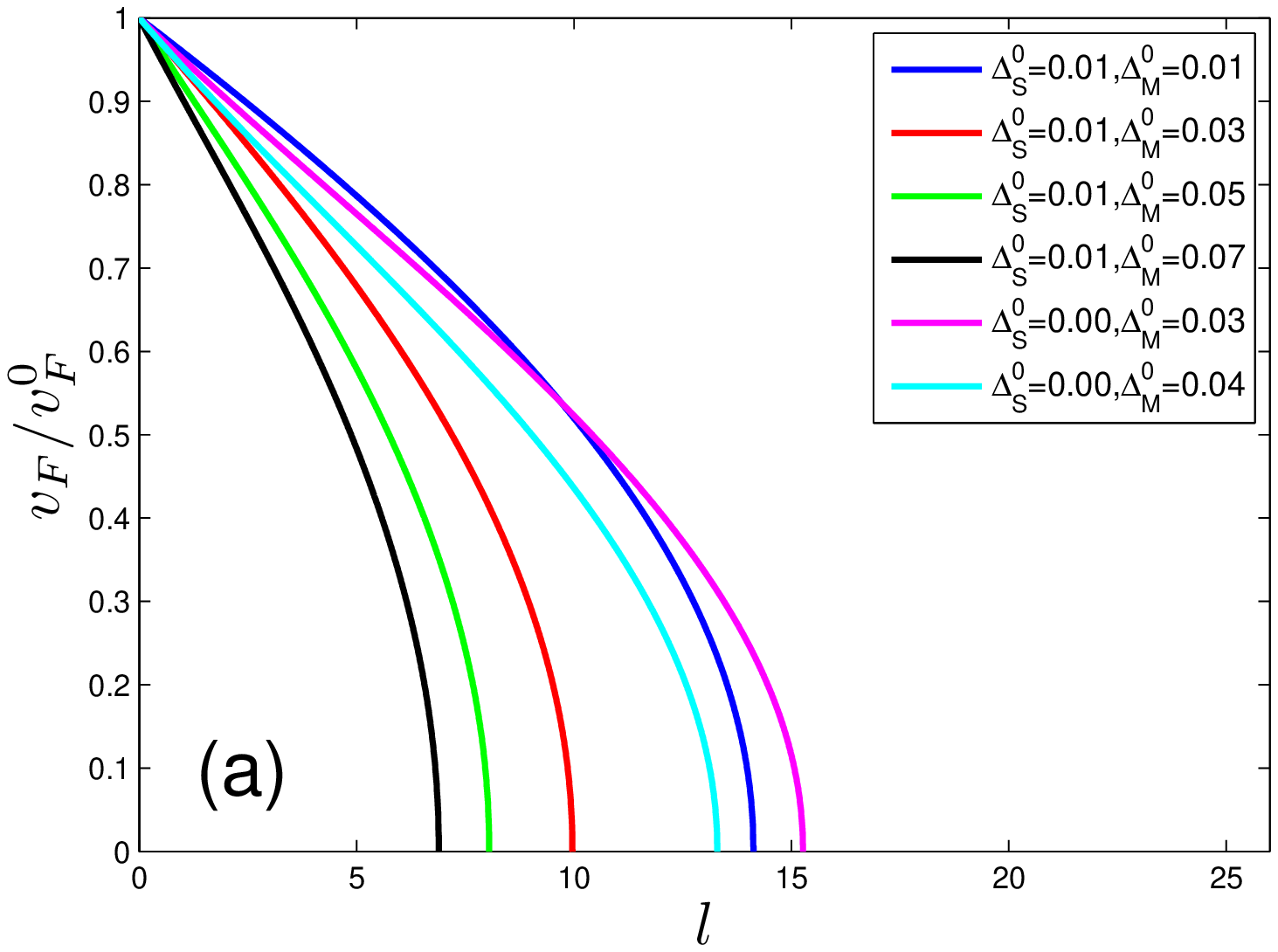}
\includegraphics[width=2.6in]{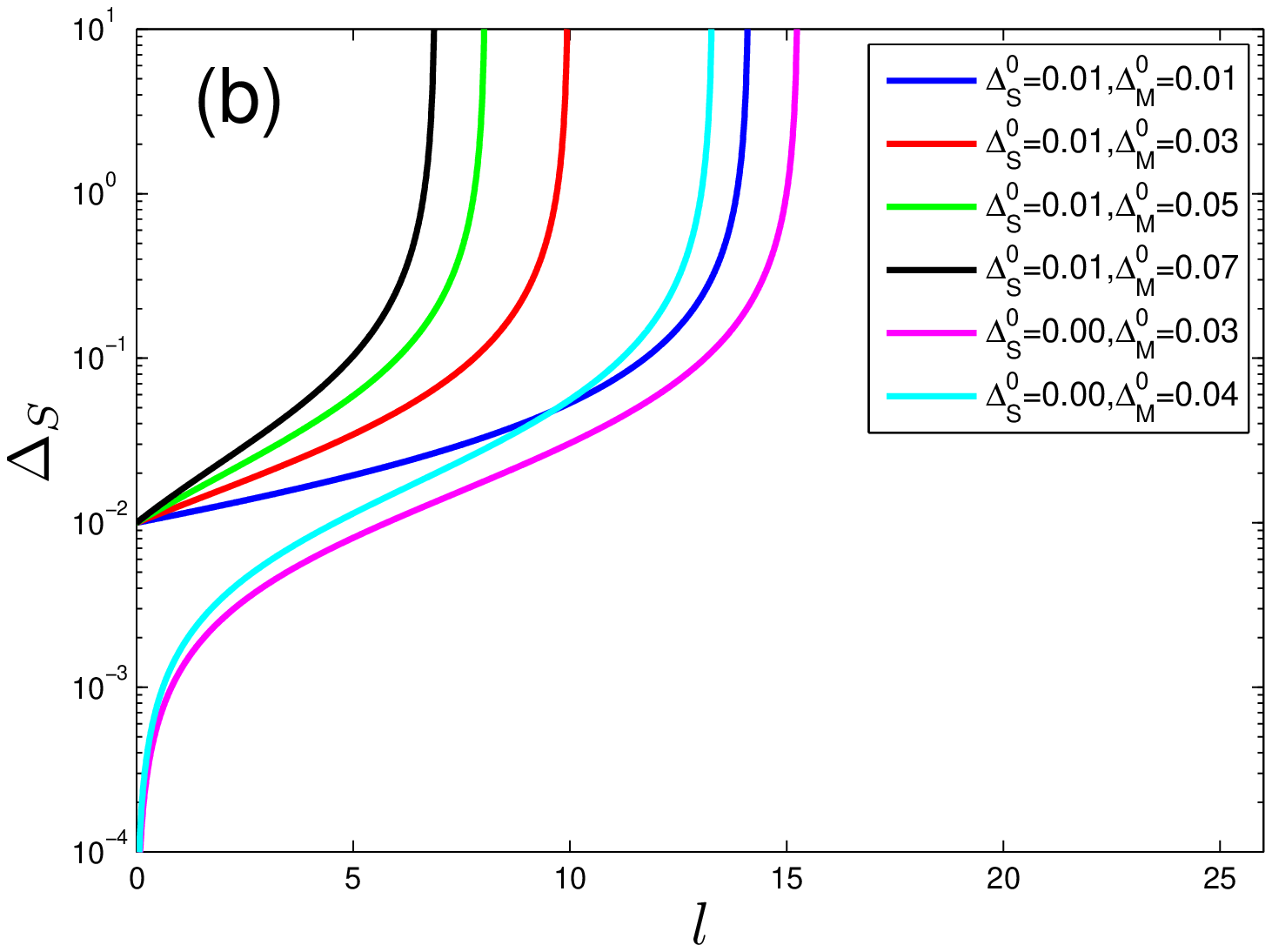}
\includegraphics[width=2.6in]{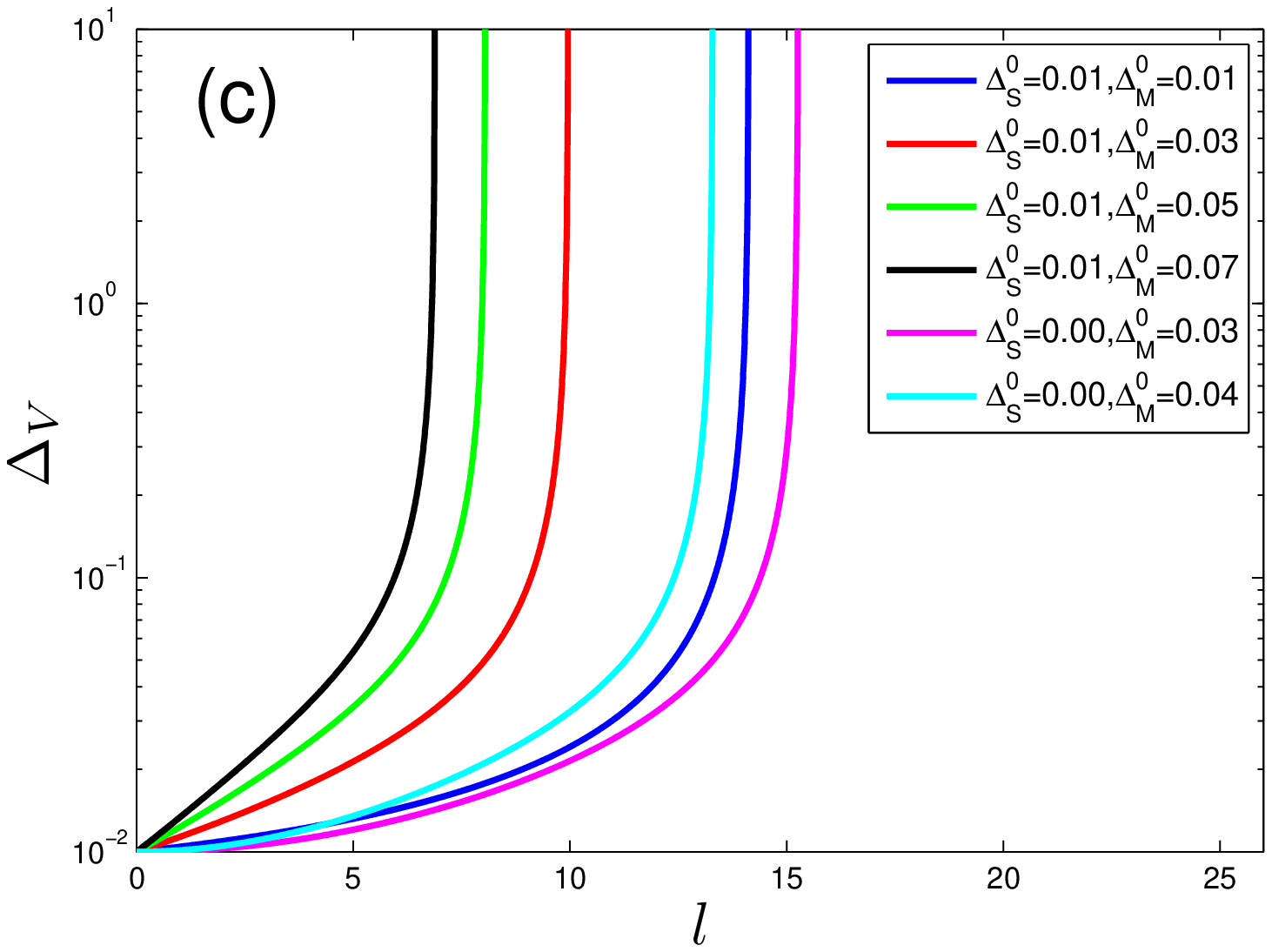}
\includegraphics[width=2.6in]{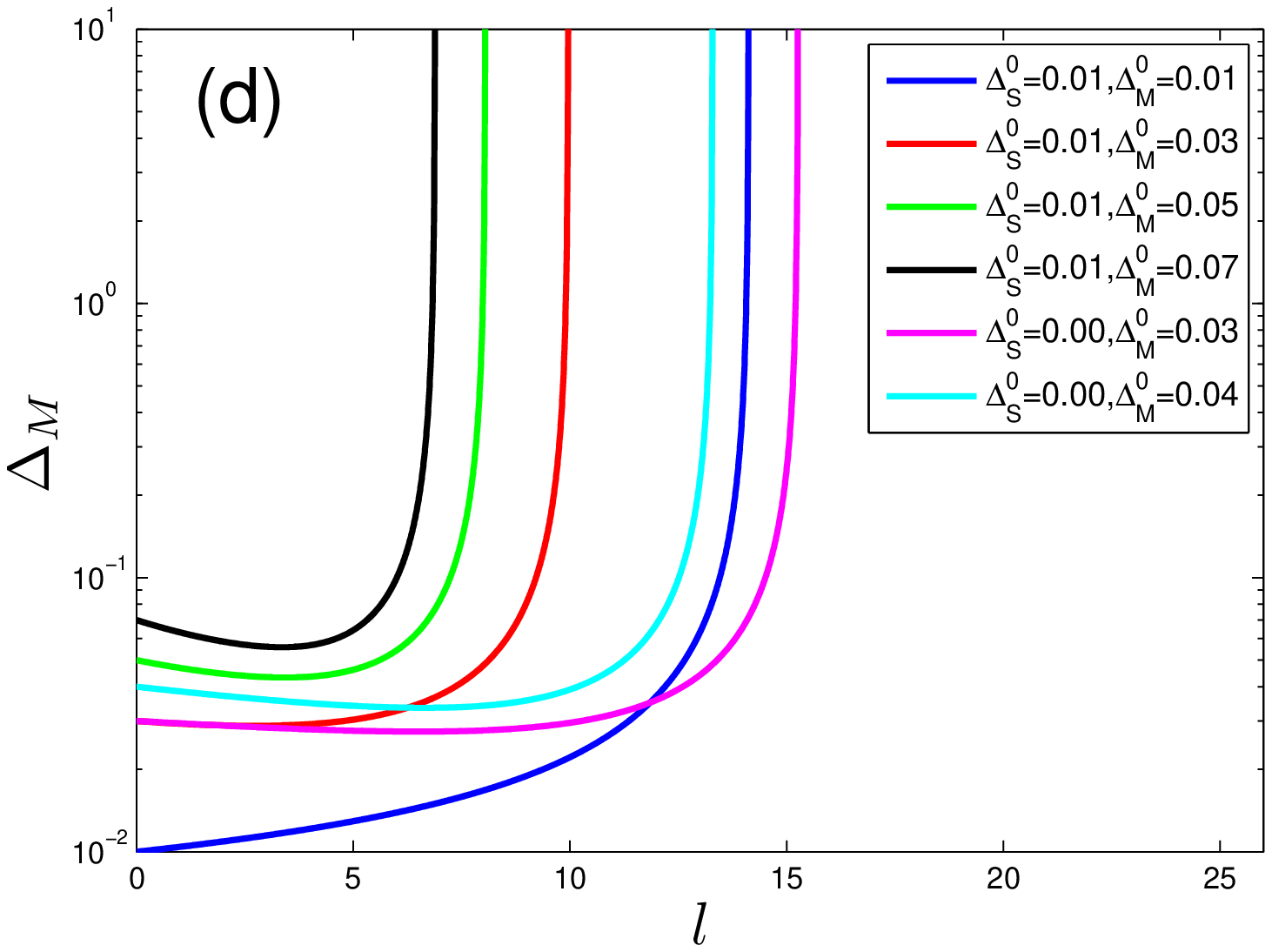}
\caption{Flowing behavior of the parameters $v_{F}$, $\Delta_{S}$,
$\Delta_{V}$, and $\Delta_{M}$ in case the system contains
multi-type of disorders, where $\Delta_{V}^{0} = 0.01$. The runaway
behaviors of $\Delta_{S}$, $\Delta_{V}$, and $\Delta_{M}$ are not
quantitatively reliable, but should be regarded as a signature of
the dominance of random chemical potential in the low-energy
region.} \label{Fig:MuxltiAllVRG}
\end{figure*}

\subsection{Coexistence of two or three types of disorder}\label{Sec_multi_dis}

In this subsection, we consider the mutual influence between
distinct types of disorder. The full set of RG equations are already
given by Eqs.~(\ref{Eq:MultiRGVF})-(\ref{Eq:MultiRGu}), and can be
solved self-consistently. For different initial conditions, the
solutions are presented in Fig.\ref{Fig:MuxltiAllVRG}.

As shown in Fig.~\ref{Fig:MuxltiAllVRG}(a), the fermion velocity
$v_{F}$ always flows to zero at some particular energy scale.
According to Fig.~\ref{Fig:MuxltiAllVRG}(b) and (c), $\Delta_{S}$
and $\Delta_{V}$ increase with lowering energy scale monotonously,
and appear to diverge at a finite energy scale. From
Fig.~\ref{Fig:MuxltiAllVRG}(d), we see that $\Delta_{M}$ increases
with lowering energy scale monotonously if $\Delta_{M}^{0} <
2\Delta_{V}^{0}$, but displays a non-monotonic dependence on energy
scale if $\Delta_{M}^{0} > 2\Delta_{V}^{0}$. An important fact is
that, once more than one types of disorder coexist, the three
parameters $\Delta_{S}$, $\Delta_{M}$, an $\Delta_{V}$ all flow to
strong couplings inevitably at low energies. This clearly informs us
that distinct types of disorder are strongly correlated with each
other, as can be seen from Fig.~\ref{Fig:MuxltiAllVRG}.

When the disorder strength flows to the strong coupling regime, it
is usually believed that such behavior leads to a finite zero-energy
DOS $\rho(0) \neq 0$ and a finite scattering rate, which drives the
Dirac fermions to enter into a diffusive phase. As just discussed,
the perturbative RG expansion method becomes out of control. In this
case, one may attempt to study the effects of disorder on
superconductivity by carrying out a mean-field analysis
\cite{Sondhi13}. For instance, it would be possible to derive the
superconducting gap equation after properly taking into account the
impact of disorder. This approach has been extensively investigated
in conventional dirty superconductors, and naturally leads to
Anderson theorem. However, 2D Dirac semimetals exhibit interesting
new features comparing to 3D ordinary metals, and one needs to be
careful in the derivation of gap equation. This issue will be
discussed in more detail in the next section.

\subsection{Effects of ZS and ZS' diagrams}\label{Sec_ZS_ZS_prime}

We have elucidated in Sec.~\ref{Sec_clean} that the contributions of
ZS and ZS' diagrams can be approximately neglected in the low-energy
regime. That consideration applies only to the clean limit. We now
incorporate the contributions of ZS and ZS' diagrams into the RG
equations and estimate their effects in the presence of disorders.

After including ZS and ZS' diagrams, the RG equations of $v_{F}$ and
$v_{\Gamma_{i}}$ are still given by
Eqs.~(\ref{Eq:MultiRGVF})-(\ref{Eq:MultiRGVGamma3}), but the RG
equation for $u$ should be modified. If the system contains only
random mass, we have
\begin{eqnarray}
\frac{du}{dl} = -\left[1 + \Delta_{M} + \frac{u[1 + 2(1 -
f'_Q)]}{8\pi v_F}\right]u, \nonumber \label{Eq:MultiZSRGu}
\end{eqnarray}
where $f'_Q \equiv \frac{2 - 2(1 - \delta)^{\frac{3}{2}}}{3\delta}$.
If the system contains only random gauge potential, one simply
replaces $\Delta_M$ with $2\Delta_{V}$. We only consider the case
that random mass or random gauge potential exists alone, since
otherwise the perturbative RG method would be out of control. The
numerical RG solutions suggest that a large value of $\delta$ favors
superconductivity, whereas a small $\delta$ disfavors
superconductivity. The influence of ZS and ZS' diagrams are
determined by the value of transferred momenta $Q$. Disorder effects
are dominant for small $Q$, but ZS and ZS' contributions become
prevailing for large $Q$. In the RG analysis, we eventually would
take $Q$ to vanish in the lowest-energy limit, which corresponds to
$\delta \rightarrow 0$. In this limit, the contributions of ZS and
ZS' diagrams become progressively unimportant.

\section{Further discussions about disorder effects}
\label{Sec_discuss}

The impact of disorder on superconductivity has been extensively
studied for nearly six decades, in the contexts of both conventional
metal superconductors and various unconventional superconductors
\cite{Anderson1959JPCS, Gorkov08, LeeRMP1985, Balatsky06}. In
particular, Anderson \cite{Anderson1959JPCS} pointed out that the
time-reversed exact eigenstates of electrons can still pair up in
disordered metals. For an $s$-wave superconductor, one can show via
gap equation calculations \cite{Gorkov08} that weak non-magnetic
disorders do not affect the superconducting gap $m$ and the critical
temperature $T_c$. In the case of 2D disordered Dirac semimetals, it
should be still possible for the exact eigenstates of Dirac fermions
related by time-reversal symmetry to form Cooper pairs. However, the
magnitude of gap $m$ and $T_c$ might be influenced by disorder
\cite{Sondhi13}.

When a 2D Dirac semimetal contains a weak random mass or gauge
potential disorder, the time-reversal symmetry is broken
\cite{Ludwig1994}. The Anderson theorem thus becomes invalid and
cannot be used to determine the fate of superconductivity. We
studied this issue by means of perturbative RG method in the last
section, and showed that the disorder strength either flows to zero
or is fixed at a small constant in the low-energy region. In both
cases, the perturbative expansion is under control. It can also be
deduced that the Dirac fermions remain extended. From the RG results
presented in Sec.V, we know that either random mass or random gauge
potential leads to suppression of superconductivity.

Random chemical potential is quite different since it preserves the
time-reversal symmetry. Different from conventional $s$-wave metal
superconductors, both the superconducting gap $m$ and
temperature $T_c$ can be modified by random chemical potential. As
mentioned in the last section, perturbative RG cannot be used to
address this issue. A promising alternative is to perform a detailed
gap equation analysis.

For conventional $s$-wave dirty superconductors, the impact of
disorder on superconductivity can be investigated by using the AG
diagrammatic approach \cite{Gorkov08}. This approach works as
follows. When a conventional $s$-wave superconductor contains weak
non-magnetic disorder, which exists as a random chemical potential,
one can derive the following gap equation \cite{Gorkov08}
\begin{eqnarray}
m = \frac{u}{4}T\sum_{\omega_{n}}\int
\frac{d^{d}\mathbf{k}}{(2\pi)^{d}}
\frac{A_{3}m}{A_{1}^{2}\omega_{n}^{2} +
\xi_{\mathbf{k}}^{2}+A_{3}^{2}m^{2}},
\end{eqnarray}
where $u$ is the BCS coupling constant, $m$ is the
superconducting gap, $A_1$ is the renormalization factor of fermion
energy, and $A_{3}$ is the renormalization factor of gap. In the
clean limit, $A_1 = A_3 = 0$. To integrate over $\mathbf{k}$, one
usually needs to make an essential assumption that the Fermi surface
is large enough such that the influence of disorder on the
low-energy DOS can be neglected. In a 3D ordinary metal, this
assumption is certainly satisfied, which allows one to employ the
approximation
\begin{eqnarray}
\int \frac{d^3\mathbf{k}}{(2\pi)^{3}}\rightarrow \int d\xi
\rho(\xi)\rightarrow \rho(0)\int d\xi,\label{Eq:ApproximationKey}
\end{eqnarray}
This then directly leads to the following gap equation
\begin{eqnarray}
m &=& \frac{u}{4}T\sum_{\omega_{n}}\rho(0)\int d\xi
\frac{A_{3}m}{A_{1}^{2}\omega_{n}^{2} +
\xi^{2}+A_{3}^{2}m^{2}} \nonumber \\
&=& \pi \frac{u}{4}T \rho(0)\sum_{\omega_{n}}
\frac{A_{3}m}{\sqrt{A_{1}^{2}\omega_{n}^{2} + A_{3}^{2}m^{2}}}.
\end{eqnarray}
Within the AG formalism, one can prove that $A_{1} = A_{3}$, which
immediately indicates that the disorder-induced renormalization
factors $A_1$ and $A_3$ cancel each other exactly. Now the gap
equation is further simplified to
\begin{eqnarray}
m = \pi \frac{u}{4}T \rho(0) \sum_{\omega_{n}}
\frac{m}{\sqrt{\omega_{n}^{2}+m^{2}}}. \label{Eq:GapDisordr}
\end{eqnarray}
This equation has precisely the same expression as that obtained in
a perfectly clean $s$-wave superconductor \cite{Gorkov08}, which
means that the superconducting gap is independent of weak random
chemical potential.

Checking the computational process, we can see that the independence
of superconductivity on disorder is based on an important assumption
that weak disorder has nearly no effects on the low-energy DOS. In
case this assumption is invalid, the disorder does not disappear.
Different from 3D ordinary metals, 2D Dirac semimetal does not have
a large Fermi surface, but has only discrete band-touching Dirac
points. Near the Dirac points, the low-energy DOS of Dirac fermions
depends on energy as $\rho(\omega) \propto |\omega|$, in the clean
limit. In particular, $\rho(0) = 0$ at the Fermi level. Once random
chemical potential is added to the system, its effective strength
increases monotonously in the low-energy region, which is often
interpreted as the emergence of a disorder-controlled diffusive
state \cite{Ludwig1994, Altland02}. A characteristic property of
such a diffusive state is the generation of a finite zero-energy
DOS, namely $\rho(0) \neq 0$. Since the zero-energy DOS is
significantly altered by random chemical potential, both the gap
$\Delta$ and $T_c$ are disorder dependent. To address this issue, we
now apply the AG formalism to examine the impact of random chemical
potential on superconductivity in 2D Dirac semimetal. After carrying
out length calculations, we obtain two self-consistently coupled
equations
\begin{eqnarray}
A &=& \!\!1 + \zeta A\ln\left(1 + \frac{1}{A^{2}\omega^{2} +
A^{2}m^{2}}\right), \label{Eq:GapEqAG1}\\
1 &=& \!\!\frac{1}{2\pi}\frac{u}{u_{c}}\int_{-\infty}^{+\infty} \!\!d\omega
A\ln\left(1+\frac{1}{A^{2}\omega^{2}+A^{2}m^{2}}\right),
\label{Eq:GapEqAG2}
\end{eqnarray}
where $\zeta=\frac{\Delta_{V}^{2}}{4\pi v_{F}^{2}}$. Here, the
renormalization factor $A_1$ and $A_3$ are still equal, and can be
universally represented by $A$ for simplicity. In the derivation of
these two equations, one cannot adopt the approximation given by
Eq.~(\ref{Eq:ApproximationKey}), but should integrate over momentum
straightforwardly. It is apparent that the factor $A$ does not
disappear and indeed satisfies two self-consistently coupled
equations. The quantities $A$ and $m$ should be computed by solving
these two equations simultaneously. Because the factor $A$ is
induced by random chemical potential, the gap is definitely not
independent of disorder.

\begin{figure}
\center
\includegraphics[width=2.7in]{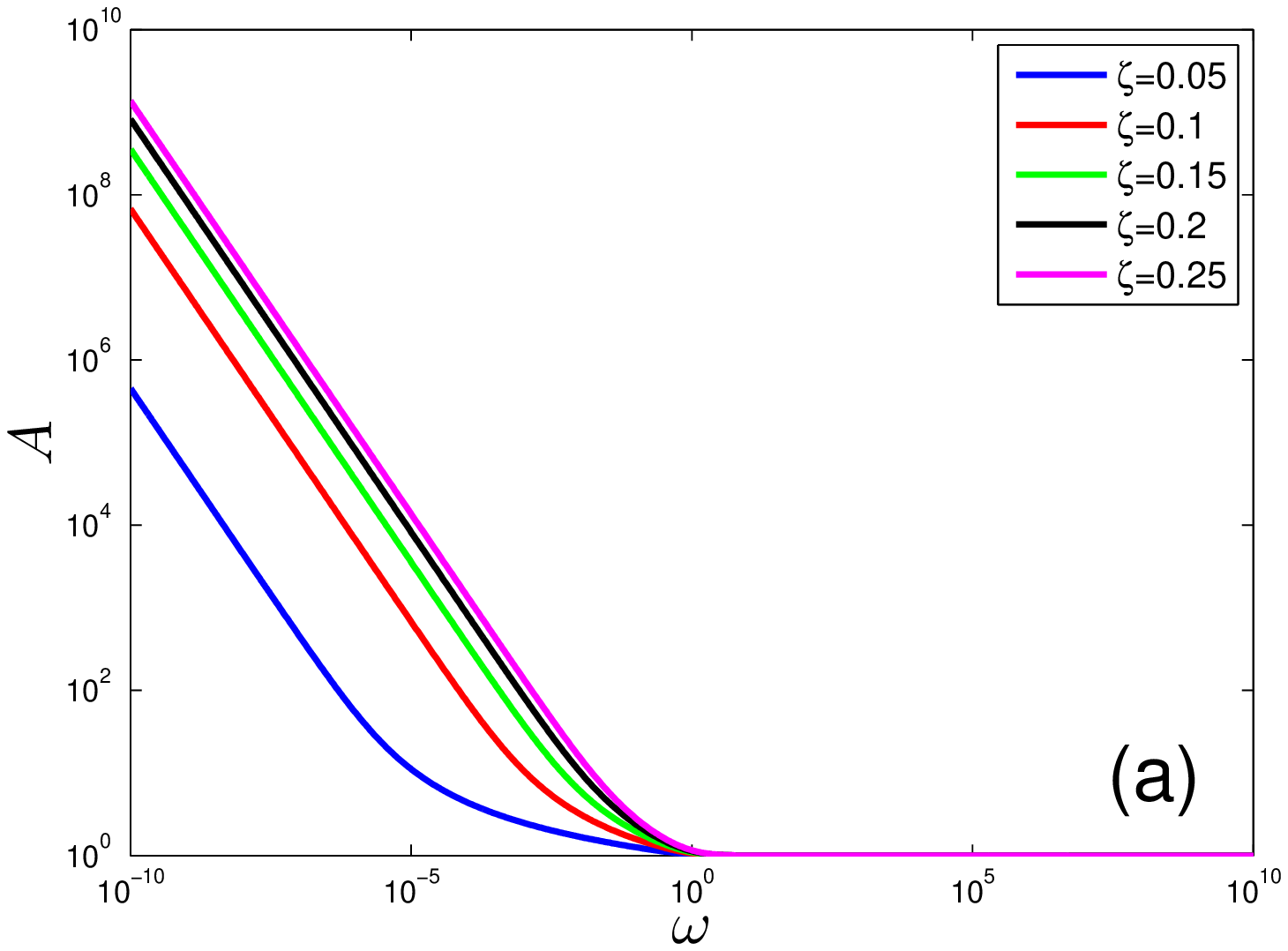}
\includegraphics[width=2.7in]{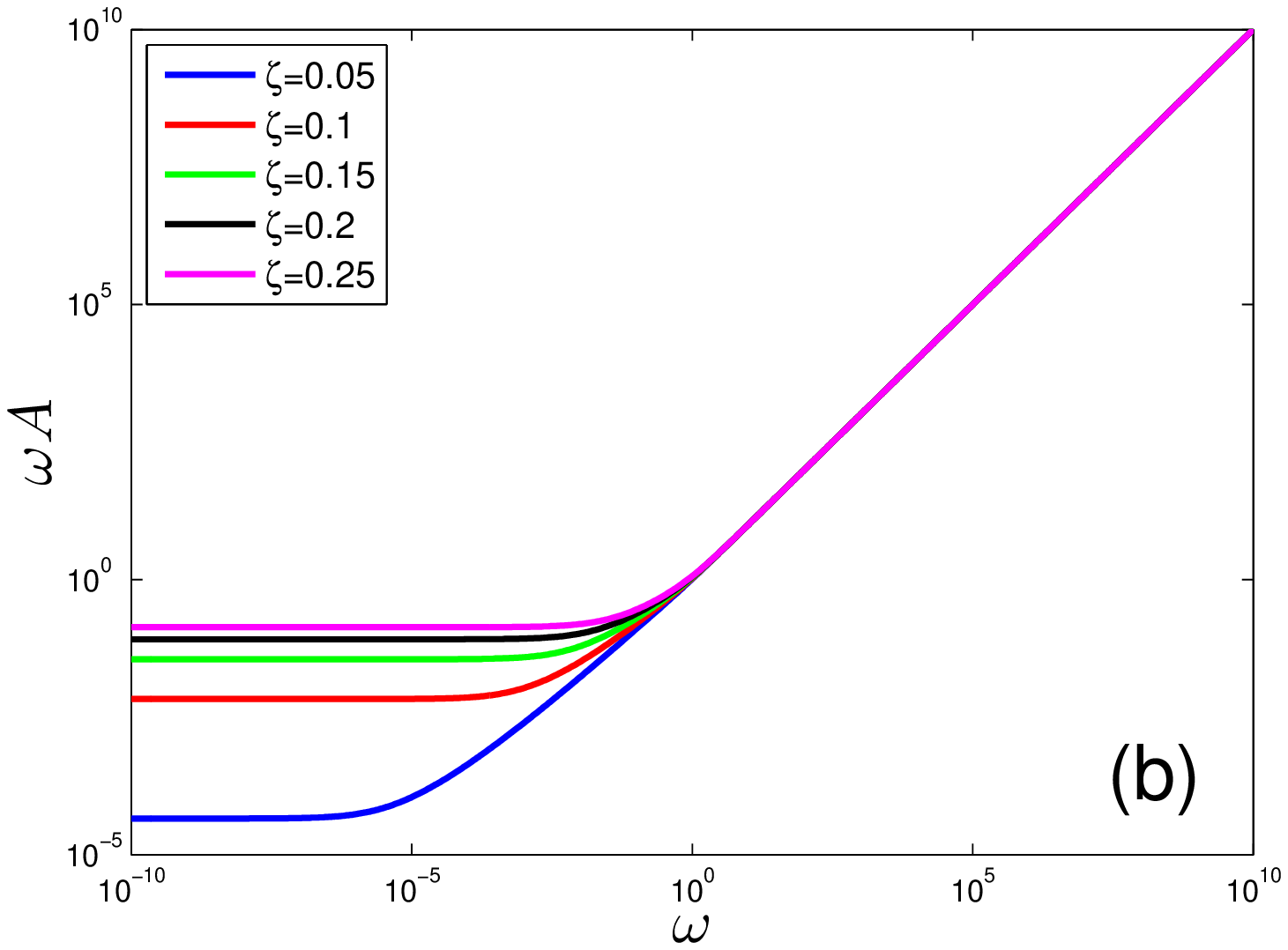}
\caption{(a) Dependence of $A$ on $\omega$ at different values of
$\zeta$. (b) Dependence of $\omega A$ on $\omega$ at different
values of $\zeta$. \label{Fig:AFun}}
\end{figure}

Now let us solve the two coupled equations (81) and (82). As the
system approaches the semimetal-superconductor QCP, i.e., $u
\rightarrow u_{c}$, the superconducting gap vanishes continuously,
and these equations can be simplified to
\begin{eqnarray}
A &=& 1 + \zeta A\ln\left(1+\frac{1}{A^{2}\omega^{2}}\right),
\label{Eq:AFunZeroGap} \\
1 &=& \frac{1}{2\pi}\frac{u}{u_{c}}\int_{-\infty}^{+\infty}d\omega A
\ln\left(1+\frac{1}{A^{2}\omega^{2}}\right).\label{Eq:EqForgc}
\end{eqnarray}
The numerical solutions of these equations are depicted in
Fig.~\ref{Fig:AFun}, which shows that $\omega A$ approaches to some
constant $\gamma$ in the zero-energy limit. The value of $\gamma$ is
determined by the strength of random chemical potential. As $\omega$
decreases from the scale set by $\gamma$, $\omega A$ becomes a
constant. If $\omega$ increases from $\gamma$, $A \rightarrow 1$.
Thus, the asymptotic behavior of $A$ is approximately given by
\begin{eqnarray}
A\sim\left\{
\begin{array}{lll}
\frac{\gamma}{|\omega|} & \mathrm{if} & |\omega| \ll \gamma,
\\
1 &\mathrm{if} & |\omega| \gg \gamma.
\end{array}\right.
\end{eqnarray}
According to this behavior, we find that the integration over
$\omega$ in Eq.~(\ref{Eq:EqForgc}) is divergent, which indicates
that
\begin{eqnarray}
u_{c} \rightarrow 0. \nonumber
\end{eqnarray}
This result means that an arbitrarily weak attraction is able to
cause BCS pairing in the presence of random scalar potential, which
can be considered as a disorder-induced enhancement of
superconductivity \cite{Sondhi13}.

\begin{figure}
\center
\includegraphics[width=2.7in]{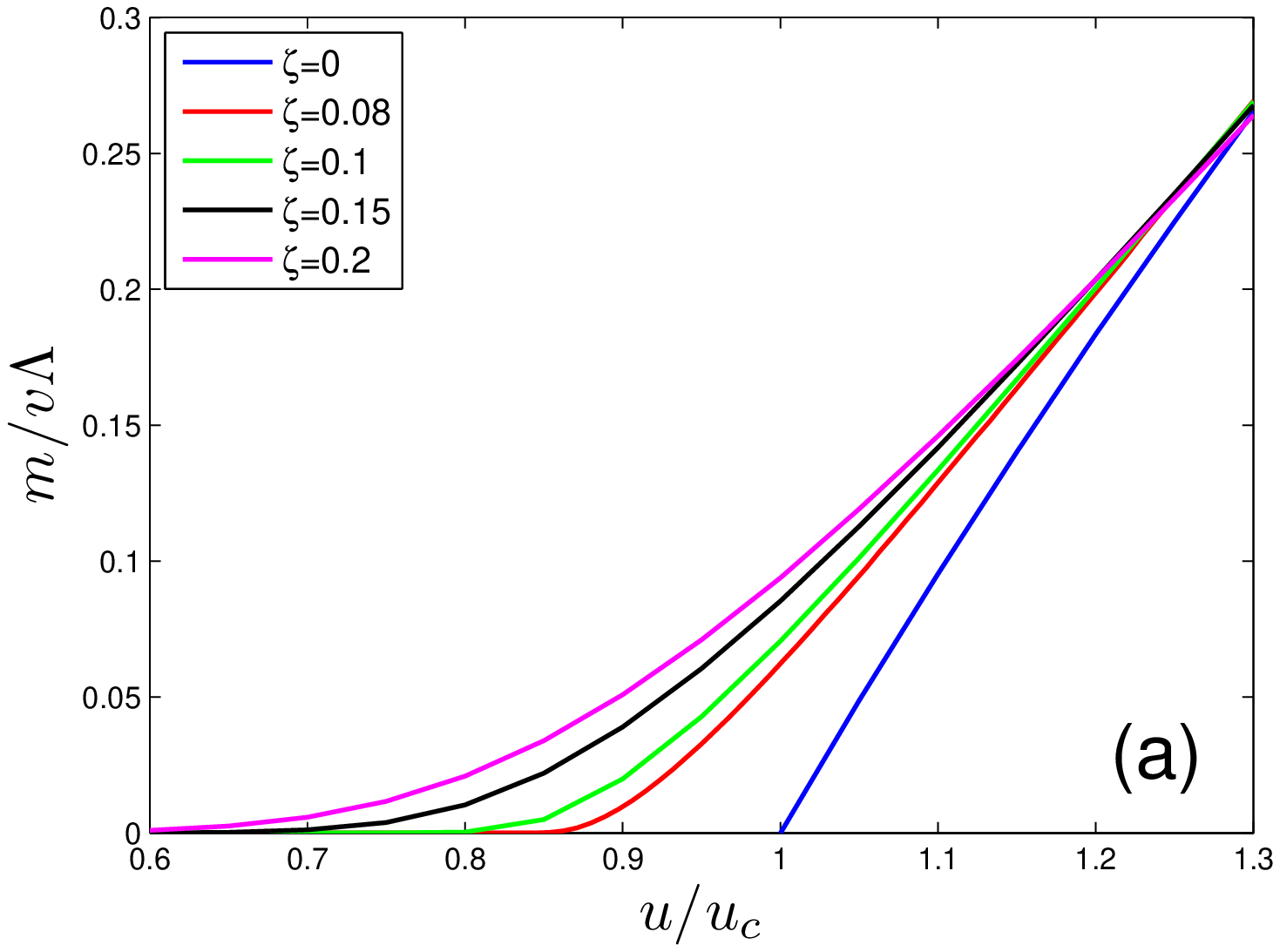}
\includegraphics[width=2.66in]{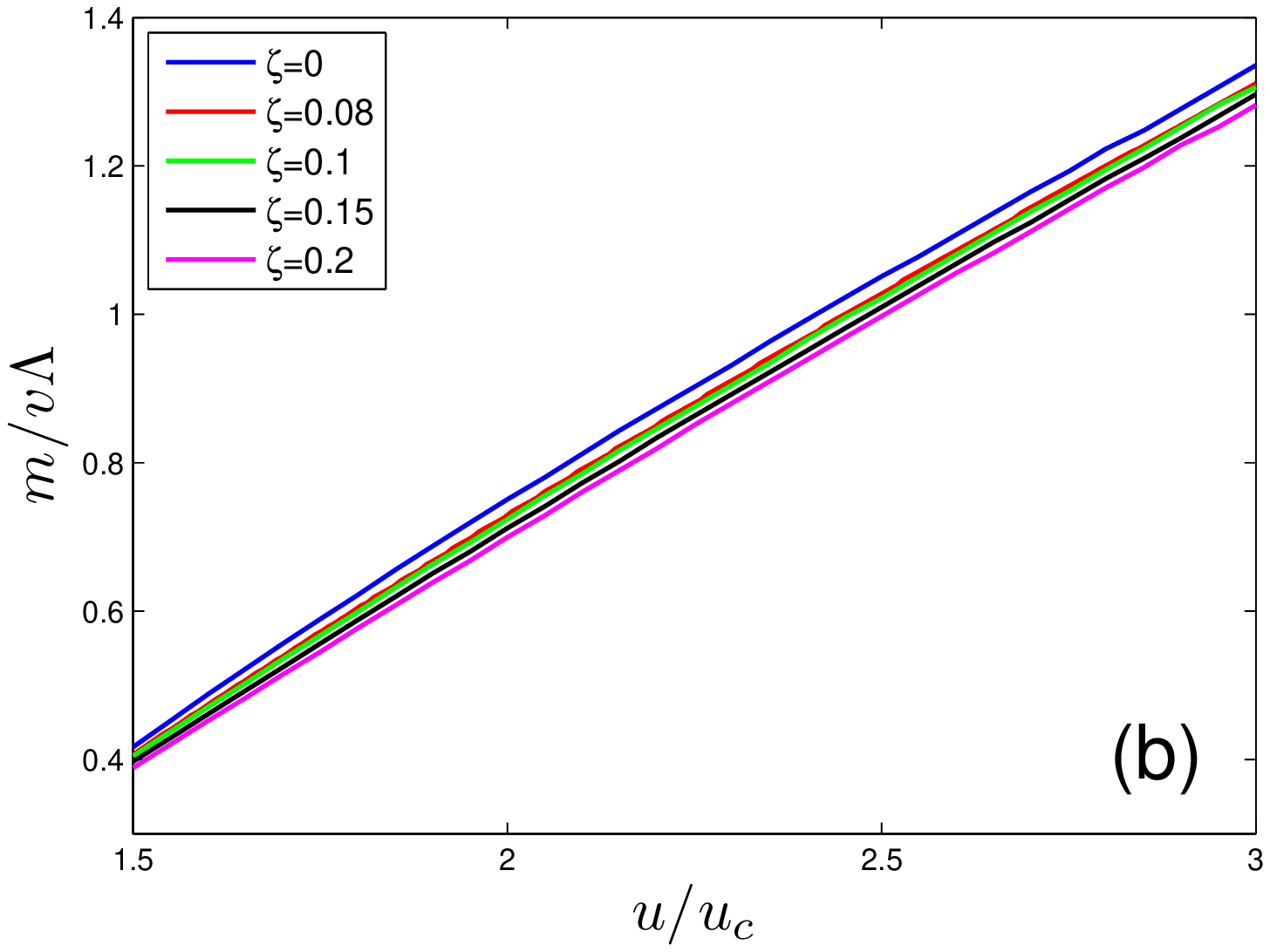}
\caption{Dependence of gap $m$ on parameter $u$ at different values
of $\zeta$. \label{Fig:Gap}}
\end{figure}

We show the dependence of gap $m$ on $u$ at different values of
disorder strength $\zeta$ in Fig.~\ref{Fig:Gap}. We find that for
small $u$, the gap $m$ is enhanced by disorder, which is shown in
Fig.~\ref{Fig:Gap}(a). However, when $u$ is larger than some
critical value, the gap $m$ is suppressed by disorder, as can be
seen from Fig.~\ref{Fig:Gap}(b). This result is qualitatively
consistent with that of Potirniche \emph{et al.} \cite{Sondhi14},
who studied the problem by solving self-consistent Bogoliubov-de
Gennes equations.

However, we should warn that the coupled equations
(\ref{Eq:GapEqAG1}) and (\ref{Eq:GapEqAG2}) may still be inadequate.
In 3D ordinary metals, the validity of AG treatment is based on an
assumption that the vertex corrections are unimportant. This
assumption is justified once the inequality $k_F \lambda \gg 1$ is
satisfied \cite{Gorkov08}. Dirac semimetals are quite different from
ordinary metals since $k_F \rightarrow 0$, hence we can no longer
utilize the condition $k_F \lambda \gg 1$ to judge whether the AG
formalism is applicable or not. The vertex corrections may play an
important role in the present system. It is an interesting task to
generalize the AG approach by incorporating the vertex corrections
in a self-consistent way, and examine the impact of random chemical
potential on superconductivity. This will be carried out in the
future work.

If random chemical potential coexists with random mass or random
gauge potential, or if all the three types of disorder are present,
the disorder strength parameters flow inevitably to strong couplings
\cite{Foster12}, and the perturbative RG method is unable to provide
a reliable tool for the determination of the fate of
superconductivity in the low-energy regime. This issue might also be
addressed by a proper generation of the AG diagrammatic approach. At
first glance, this situation seems to be quite similar to the case
in which random chemical potential exists alone. However, they are
actually different. The time-reversal symmetry is preserved when the
system contains only random chemical potential, but is explicitly
broken once random mass and/or random gauge potential exist
\cite{Evers08, Mirlin2006PRB, Foster12}. It would be interesting to
study whether such symmetry breaking has a remarkable impact on
superconductivity.

\section{Summary and discussion}\label{Sec_summary}

In this paper, we have studied the interplay between an effective
BCS-type interaction and fermion-disorder coupling by performing a
RG analysis. Our main finding is that random mass and random gauge
potential both lead to certain amount of increment of the critical
BCS coupling $|u_{c}|$, which makes the onset of superconductivity
more unlikely since a stronger net attraction is required to form
Cooper pairs comparing to the clean case. In addition to the
suppression of superconductivity, disorders have other drastic
effects on the low-energy behaviors of Dirac fermions. At the new
critical value $|u_{c}'|$, which is lager than $|u_{c}|$ obtained in
the clean limit, the system would undergo a quantum phase transition
between a superconducting phase and: (a) a MFL-like phase in the
case of random mass; (b) a NFL in the case of random gauge
potential. It is interesting to study the critical properties of
these QCPs, and examine whether an effective supersymmetry emerges
in the low-energy regime.

In the case of random chemical potential, our RG analysis show that
the effective disorder parameter $\Delta_{S}$ grows monotonously as
energy is lowered. This indicates that such type of disorder plays a
significant role at low energies, but also signals the breakdown of
perturbative RG method. We have investigated the impact of random
chemical potential on superconductivity by carrying out a
straightforward AG analysis and compared to previous pertinent
works. We then have demonstrated that such simple AG analysis might
be insufficient to get a reliable conclusion and that the AG
diagrammatic approach needs to be improved to include the vertex
corrections self-consistently.

We also have considered the mutual influence between different types
of disorder. In case more than one types of disorder coexist, all
three types of disorder should be present and their effective
strength parameters all flow to strong couplings \cite{Foster12}.
Given that perturbative RG approach become inapplicable, other
efficient theoretic techniques are urgently needed to handle this
complicated problem.

We then briefly discuss the case of doped 2D Dirac semimetal. For a
2D Dirac semimetal defined at a finite chemical potential $\mu$,
previous mean-field studies have found that an arbitrarily small
attraction suffices to induce Cooper pairing \cite{Uchoa05,
Kopnin08}. In light of these studies, we expect that the same
conclusion should be reproduced by the RG method. In particular, the
RG equation would be the same as Eq.~(\ref{Eq_u_c}). If this is the
case, the critical BCS coupling should vanish: $u_c = 0$. However,
we need to emphasize that the problem of Cooper pairing becomes
highly nontrivial when a 2D Dirac semimetal is doped. For the
surface state of 3D topological insulator, the paring symmetry is
$s$-wave at zero doping. At a finite $\mu$, the $s$-wave gap will
mix with a $p_{x}+ip_{y}$-wave gap, although the $p_{x}+ip_{y}$
component is small if $\mu$ is not very large \cite{Sondhi13}. In
doped graphene, the possible pairing symmetry is more complicated
due to the presence of several valleys of Dirac fermions. Moreover,
when graphene is doped to the vicinity of the Van Hove singularity,
a $p$-wave chiral superconductivity may emerge as the ground state.
Due to these complications, it is actually not easy to make a full
RG analysis of superconductivity. Technically, the RG scheme used at
$\mu = 0$ cannot be simply applied to the case of $\mu \neq 0$. In
the former case, all the associated momenta can be assumed to be
small quantities in the lowest-energy limit. However, in the latter
case, only the component $k_{\bot}=|\mathbf{k}|-k_{F}$ can be
considered as small at low energies. To study the latter case, one
should employ the RG scheme similar to that utilized in some recent
works on Cooper pairing in NFL systems \cite{Metlitski15,
Fitzptrick15, Raghu15}.

It is also interesting to make a RG analysis to study the impact of
various types of disorder on the Cooper pairing instability in other
semimetal materials, such as 3D Dirac semimetals \cite{Vafek}, 2D
semi-Dirac semimetals \cite{Yang14, Isobe16, Moon16}, and double-
and triple-Weyl semimetals \cite{Huang16, Lai15, Jian15}.

\section*{ACKNOWLEDGEMENTS}

The authors are supported by the National Natural Science Foundation
of China under Grants 11274286, 11574285, 11504360, and 11504379.
J.W. is also supported by the China Postdoctoral Science Foundation
under Grants 2015T80655 and 2014M560510, and the Fundamental
Research Funds for the Central Universities (P. R. China) under
Grant WK2030040074.

\end{document}